\documentclass[reprint,superscriptaddress,prl]{revtex4-2}

\usepackage[pdftex]{graphicx}
\graphicspath{{./pictures}{./Figures}}
\usepackage{amsmath, amssymb, bm, dcolumn}
\usepackage[T1]{fontenc}
\usepackage[utf8]{inputenc}
\usepackage[dvipsnames]{xcolor}
\usepackage{siunitx}
\usepackage{booktabs}
\usepackage{cancel}
\usepackage[colorlinks=true,linkcolor=blue, citecolor=violet]{hyperref}%
\usepackage[ruled,vlined]{algorithm2e}
\SetKwInOut{Input}{Input}
\SetKwInOut{Output}{Output}

\usepackage{rotating}

\begin{document} 

\title{Localization and coherent control of 25 nuclear spins in Silicon Carbide}
\author{Pierre Kuna}
\thanks{These two authors contributed equally}
 \affiliation{3rd Institute of Physics, IQST, and Research Center SCoPE, University of Stuttgart, Stuttgart, Germany}
\author{Erik Hesselmeier-Hüttmann}
\thanks{These two authors contributed equally}
 \affiliation{3rd Institute of Physics, IQST, and Research Center SCoPE, University of Stuttgart, Stuttgart, Germany}
\author{Phillip Schillinger}
 \affiliation{3rd Institute of Physics, IQST, and Research Center SCoPE, University of Stuttgart, Stuttgart, Germany}
\author{Felix Gloistein}
 \affiliation{3rd Institute of Physics, IQST, and Research Center SCoPE, University of Stuttgart, Stuttgart, Germany}
 \author{István Takács}
 \affiliation{Eötvös Loránd University, Egyetem tér 1-3, H-1053 Budapest, Hungary}
\affiliation{MTA–ELTE Lendület “Momentum” NewQubit Research Group, Pázmány Péter, Sétány 1/A, 1117 Budapest, Hungary}
 \author{Viktor Ivády}
 \affiliation{Eötvös Loránd University, Egyetem tér 1-3, H-1053 Budapest, Hungary}
\affiliation{MTA–ELTE Lendület “Momentum” NewQubit Research Group, Pázmány Péter, Sétány 1/A, 1117 Budapest, Hungary}
 \author{Wolfgang Knolle}
 \affiliation{Department of Sensoric Surfaces and Functional Interfaces, Leibniz-Institute of Surface Engineering (IOM), Permoserstraße 15, 04318 Leipzig, Germany}
\author{Jawad Ul-Hassan}
 \affiliation{Department of Physics, Chemistry and Biology, Linköping University, Olaus Magnus väg, 583 30 Linköping, Sweden}
\author{Jörg Wrachtrup}
 \affiliation{3rd Institute of Physics, IQST, and Research Center SCoPE, University of Stuttgart, Stuttgart, Germany}
 \affiliation{Max Planck Institute for solid state physics, Heisenbergstraße 1, 70569 Stuttgart, Germany}
\author{Vadim Vorobyov}
\email[Correspondence:]{v.vorobyov@pi3.uni-stuttgart.de}
 \affiliation{3rd Institute of Physics, IQST, and Research Center SCoPE, University of Stuttgart, Stuttgart, Germany}

\begin{abstract}
Optically addressable spin defects are excellent candidate platform for quantum sensing and quantum network.
Nuclear spins coupled to color centers naturally enable long lived quantum memories and local qubits registers. 
To fully leverage this potential precise characterization of the surrounding nuclear-spin environment augmented with refined DFT models is required. 
In this work, we report angstrom-level 3D localization of 25 nuclear spins around a single V2 center in 4H Silicon Carbide. 
Utilizing specially placed robust nuclear memory as a highly efficient readout ancilla for readout, we apply correlation based spectroscopy and by selecting multi-spin chains up to length four, we access and characterize extended nuclear spin cluster.
Using the coupling map we reconstruct their couplings to the central electron spin and neighboring nuclei.
This work paves the way towards advanced quantum register applications on Silicon Carbide platform.
\end{abstract}
\maketitle
\section{Introduction}
Optically addressable spin defects in solids have become a leading platform for quantum technologies, offering a unique combination of long spin coherence, optical addressability and compatibility with established semiconductor processes \cite{awschalom2018quantum}.
When coupled to nearby nuclear spins, they form local multiqubit registers with optical interfaces - an architecture well suited for quantum communication \cite{review_englund, van2020extending}, error corrected memories \cite{waldherr2014quantum, Taminiau_2014} and distributed quantum information processing \cite{wei2025universal, de2024thresholds, simmons2024scalable}. 

Over the past decades, the nitrogen vacancy (NV) center in diamond \cite{chu2015quantum} has served as the benchmark system for this approach. It has enabled precise control over vast multi spin registers \cite{abobeih2019atomic,Cujia_2022, van2024mapping}, high fidelity initialization and readout \cite{Robledo_2011, neumann2010single, zahedian2023readout} and deterministic spin-photon entanglement for elementary quantum networks \cite{togan2010quantum, kalb2017entanglement, tchebotareva2019entanglement, bernien2013heralded, pompili2021realization, javadzade2025efficient, stolk2024metropolitan}. 
Yet, the NV system faces intrinsic limitations in scaling: its optical transitions have low Debye-Waller factor and limited spectral stability in nanophotonic structures, posing challenges for integration \cite{orphal2023optically, hausmann2013coupling, wolters2013measurement}.

To overcome this constraints, significant effort has been devoted to Group-IV centers in diamond (SiV$^-$, SnV$^-$,  GeV$^-$, PbV$^-$) whose inversion symmetry offers superior optical stability \cite{M_ller_2014, sipahigil2016integrated, Pingault_2017, siyushev2017optical, thiering2018ab}. 
However these systems introduce new challenges: SiV and GeV require millikelvin operation to preserve spin coherence \cite{PhysRevB.109.085414}, while  heavier group IV centers suffer from weak microwave drivability \cite{karapatzakis2024microwave, pieplow2024efficient, rosenthal2023microwave}.
Additionally, due to their $S=1/2$ electronic ground states, group IV offer a limited access to larger nuclear registers \cite{zahedian2024blueprint, stas2022robust, Beukers2025, resch2025high, grimm2025coherent}. 
Together, these motivates the search for an alternative defect platform that combines stable optical interface with robust spin coherence, efficient spin manipulation and access to large, addressable nuclear environment.

4H Silicon Carbide (4H-SiC) has recently emerged as a compelling material that satisfies these criteria \cite{castelletto2020silicon, son2020developing, xu2021silicon}. 
It is technologically mature semiconductor with established nanofabrication process, and its wide bandgap hosts several optically stable color centers, including silicon vacancy. 
Rapid progress has demonstrated spin photon entanglement \cite{fang2024experimental}, efficient single shot readout \cite{Hesselmeier2024_SSR,Lai2024} nanophotonic  \cite{babin2022fabrication, lukin2023two, van2025check} and electrical integration \cite{hollendonner2025quantum, anderson2019electrical,anderson2022five,steidl2025single}, positioning Silicon Vacancy center in Silicon Carbide (V$_{\mathrm{Si}}$) as promising building block for quantum networks. 
Moreover, unlike centers in diamond, $V_{\mathrm{Si}}$ in SiC naturally offers a bispecies nuclear spin environment and an electron spin $S = 3/2$, providing access to a richer and potentially larger nuclear spin register \cite{parthasarathy2023scalable, Hesselmeier2024_SSR}.
Yet, despite these advantages, coherent control of extended spin clusters and detailed microscopic characterization of the surrounding nuclear environment have remained experimentally unexplored.

\begin{figure*}
    \centering
    \includegraphics[width=\textwidth]{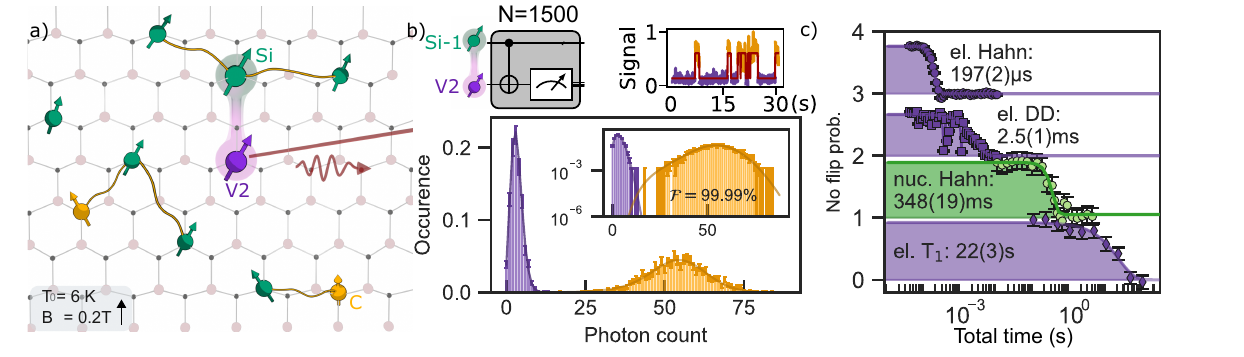}
    \caption{\textbf{a)} Schematic of V2 defect surrounded by $^{29}$Si and ${^13}$C isotopes. \textbf{b)} Single shot readout of Si1 nuclear spin qubit \textbf{c)} Electron and nuclear spin coherence $T_2$ and electron spin-lattice relaxation $T_1$ time.}
    \label{fig:fig1}
\end{figure*}

In this work we harness the coherent interactions between a cubic site silicon vacancy V$_2$ = V$_{\mathrm{Si,k}}$ central spin and its surrounding nuclear spin cluster (see Fig. \ref{fig:fig1}a) and characterize the internuclear couplings within the cluster. 
We achieve an optimised nuclear spin assisted readout with a fidelity exceeding 99 \%, enabling a high sensitivity NMR spectroscopy of more than thirty nearby nuclei. 
Utilising a combination of  electron nuclear double resonance (ENDOR), dynamical decopling based radiofrequency addressing (DDRF) and a nuclear-nuclear spin echo double resonance (SEDOR) we extract hyperfine couplings to the central electron spin and nuclear - nuclear dipole-dipole couplings within the cluster. 
An optimisation-based reconstruction algorithm allows us to find locations of twenty five spins with single site precision. 
The resulting hyperfine tensors are in quantitative agreement with density-functional theory based calculations yielding a validated microscopic model of the V$_2$ center.

These results establish a new state of the art in nuclear spin control within SiC and demonstrate all essential components of a local multi-qubit register: high-fidelity readout, precise nuclear-spin addressing, characterization of internuclear couplings and quantitative atomic reconstruction - capabilities that, until now, have only been realized within a single class of solid-state defect systems.
By achieving comparable functionality in SiC—an optically advanced and technologically mature semiconductor—we extend register-based approaches to a distinct and scalable material platform. 
This advances SiC color centers toward their full potential as robust nodes for quantum networking, long-lived quantum memories, and distributed quantum information processing using local registers built around the V$_2$ center.

\section{V2 center in 4H-SiC with nearly non-flipping nuclear spin}
The basis of our spin register is formed by the central electron spin $S = 3/2$ of the V2 center in 4H-SiC depicted on Fig. \ref{fig:fig1}. 
Magnetic field of $B\sim0.2$ T is applied along the c-axis of the SiC crystal. 
The foundation of a high-fidelity register operation is the readout and initialization of the system.
To this end a reliable quantum non-demolition measurement (QND) has been realized, analogous to described in \cite{Hesselmeier2024_SSR}.
Based on the prediction of the DFT-model \cite{Hesslemeier2024_GSLAC} a $^{29}$Si nuclear spin  coupled with $A_{zz} = -4.8$ MHz is located on the symmetry axis of the V$_2$ (Fig. \ref{fig:fig1}a), unique for the k-site of the silicon vacancy.


Its symmetric axial site enable outstanding performance during repetitive readout (SSR) technique, as quantization axis does not change when the electron spin is promoted to the excited state during the readout, yielding  lower nuclear spin flip rates compared to off-axis silicon lattice sites.
This results in a prominent telegraph like process upon continuous spin readout and the flipping rates could be directly extracted from the statistics of the switching time in the time traces (Fig. \ref{fig:fig1}b).
The resulting flip rates for the bright and dark states are 0.9(2)\,Hz and 0.18(2)\,Hz, respectively (see Supplementary Materials).
A low flipping rate allows for extended interrogation of the system under readout and yields a readout fidelity of 99.99\% under N = 1500 repetitive readout cycles.
The complete readout sequence requires approximately 7\,ms, which is negligible compared to the timescales of the sensing and quantum control sequences employed in this work.
\begin{figure*}
    \centering
    \includegraphics[width=\textwidth]{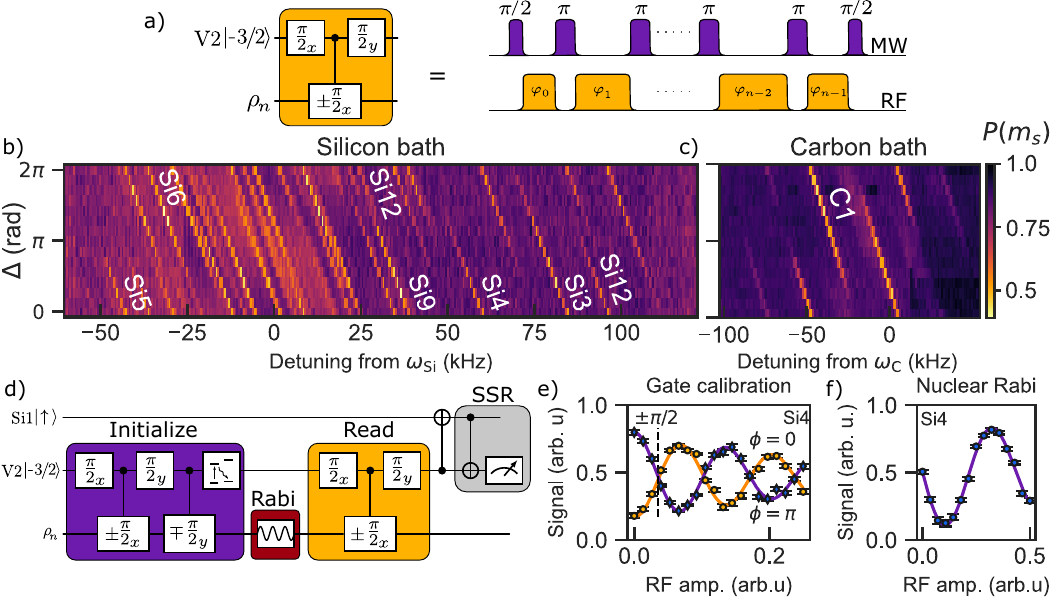}
    \caption{\textbf{a)} Quantum circuit representation and pulse-train diagram of the DDRF gate. \textbf{b)} Scan of the silicon bath. Multiple distinct lines indicate possible control over several nuclear spins. Si3 and Si4 resonances are from the $m_s=-1/2$ subspace. Two Si12 resonances can be seen due to two different electron subspaces. All other marked resonances belong to $m_s=3/2$. \textbf{c)} Scan of the carbon bath. Two resonances, according to two electron states are visible. \textbf{d)} Quantum circuit diagram for initialization, manipulation and readout of a nuclear spin via DDRF. \textbf{e)} Amplitude calibration of the DDRF gate. The dashed line indicates the amplitude which rotates the nuclear spin onto the equator of the Bloch sphere, resulting in an effective CNOT gate. \textbf{f)} Rabi oscillation of Si4. Recorded by the sequence depicted in \textbf{d}.}
    \label{fig:ddrf_rabi}
\end{figure*}

\section{Nuclear spin control}
To address the weakly coupled nuclear spins, we are utilising the central electron spin of the V2 center as well as the strongly coupled $^{29}$Si spin denoted as Si1. 
Coherent properties of the electron spin and nuclear spin utilised as a sensor are presented in Fig. \ref{fig:fig1}c.
The experimentally evaluated electron spin relaxation time of $T_1 = 22 (3)$ s is comparable to previously reported values \cite{simin2017locking}. 
Electron spin coherence using the Hahn-Echo scheme is $T_2^{HE} = 192 (2) \,\mathrm{\mu s}$, and can be prolonged using the XY-N and other dynamical decoupling \cite{steidl2025single}. In particular for N = 16 we obtain coherence time $T_2^{DD} = 2.5$ ms.
The nuclear spin Si1 coherence time with Hahn-Echo sequence is 348(19) ms.
The coherence time of the nuclear spin allows for spin echo double resonance sequence to be sensitive down to 1 Hz coupling strength using the XY-N decoupling sequences applied to the probing nuclear spin as will be described below. 
First we address the weakly coupled nuclear spins using the established dynamically decoupled radio-frequency gates (\textit{DDRF}) \cite{Bradley2019, Beukers2025, vanOmmen2025}.
\begin{figure*}
    \centering
    \includegraphics[width=\textwidth]{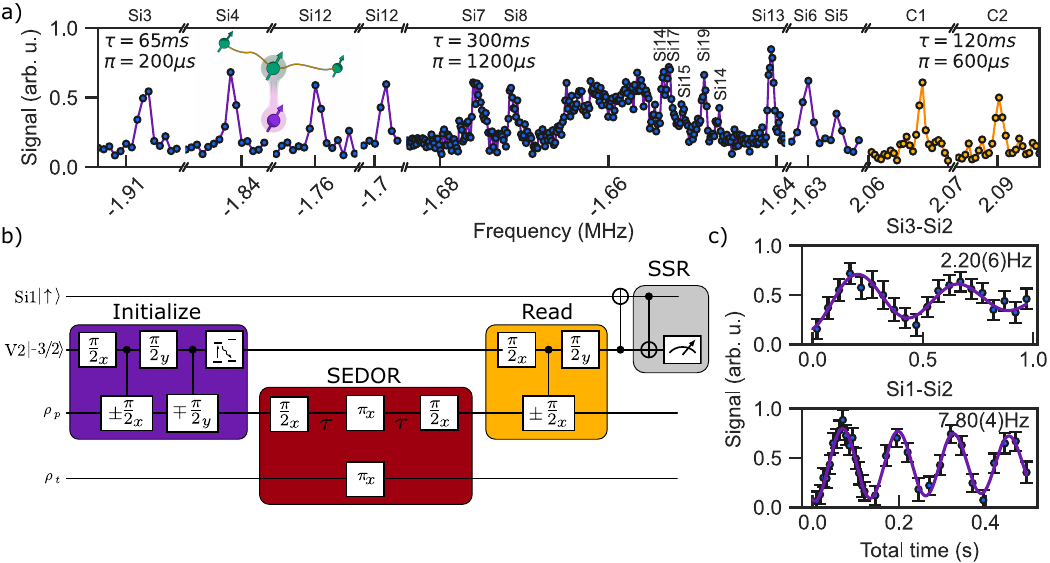}
    \caption{\textbf{a)} SEDOR measurement with Si1 being the probe spin. 14 silicion and 2 carbon spins can be observed. The contrast of each peak in the spectrum depends on the individual coupling between probe and target. Spins with small A$_\parallel$ typically are distant from Si1, thus a increased interaction time of $\tau$ is applied. The linewidth is inverse proportional to the $\pi$-pulse duration. \textbf{Inset:} Si1 is used as a probe to sense surrounding isotopes. \textbf{b)} Quantum circuit of a SEDOR measurent. The probe is initialisd either via SSR (Si1) or via DDRF (all others). c) Scan of the interaction time $\tau$ at a fixed target frequency.}
    \label{fig:sedor_spectrum}
\end{figure*}

DDRF is a state-of-the-art sequence for controlling a weakly coupled nuclear spins via radiofrequency pulses (RF) coupled to an electron spin and additionally serves to protect the electron spin coherence via dynamical decoupling. 
We adapt the sequence here for the two sub-levels of the electron spin $S=3/2$, using the $m_s = +3/2 \longleftrightarrow +1/2$ or $m_s = -3/2 \longleftrightarrow -1/2$ as the working two-level subspace.
The sequence in the $m_s = -3/2 \longleftrightarrow -1/2$ subspace runs as following (Fig. \ref{fig:ddrf_rabi}):
after initialization of the strongly coupled nuclear spin the electron spin is prepared in $m_s=-3/2$ via optical pumping and selective microwave driving \cite{nagy2019high} a Knill dynamical decoupling sequence (KDD-XY) is applied to the electron spin: a train of N evenly spaced microwave (MW) $\pi$ pulses flips the electron between the two states of the subspace (Fig. \ref{fig:ddrf_rabi}a).
The interpulse spacing of 2$\tau\approx40 \, \mu$s between the $\pi$ pulses is chosen such that it matches the Larmor frequency of the silicon spin bath.
An RF field at the nuclear resonance frequency for the $m_s = -3/2$ electron state is applied during the evolution time $2\tau$ of the DD sequence.
While the electron spin is in the state $m_s=-1/2$ the nuclear spin is off-resonant to the RF drive and picks up a phase due the precession and off-resonant driving.
Thus, after each $\pi$ pulse the phase $\varphi$ of the RF drive is updated with phase shift $\Delta$ such that all RF pulses add constructively.
Additionally a $\pi$ phase on every second RF pulse to control axis of rotation depending on the initial electron spin state,
yielding $\delta \varphi_{i,i+1} = \pi + \Delta$ (see Supplementary Material).
Hence when the resonance condition is met $\Delta  = (\omega_0 + \omega_1 -2 \omega_{rf})\cdot \tau$, the resonances corresponding to single nuclear spins are seen.
The full spectra of experimentally measured Silicon and Carbon nuclear spin bath in DDRF sequence, with varied frequency of the RF drive, the $\Delta$ phase update are presented on the Fig. \ref{fig:ddrf_rabi}b,c respectively.
Each point in the 2D map is the remaining coherence on the electron spin obtained by a phase sweep of the last microwave $\pi/2$ pulse in DDRF sequence. 
In contrast to systems with electron spin $S=1$, like the NV center in diamond or VV$^0$ in SiC, the V$_2$ center investigated here is a $S=3/2$ system and, in particular, lacks an $m_s=0$ state. 
As a result, the nuclear transition frequency of one spin in the $m_s =-3/2$ manifold can coincide with that of a different spin in the $m_s=-1/2$ manifold. 
This is seen on the example of the spectra of Si12 spin, which appears twice in the spectra, for $m_s = 3/2$ and for $m_s = 1/2$ respectively. 
For weaker coupled spins, this spectral overlap complicates selective nuclear spin addressing via DDRF in regions with high spectral crowding. 
For attributing the individual spins to the transitions in DDRF sequence and avoid ambiguity, we utilized the positions of the spin resonances obtained from ENDOR spectroscopy (see Supplementary information for details) and related them to the expected DDRF positions using numerical simulation. 
In this work the selective addressing via DDRF was applied only to nuclear spins with parallel hyperfine couplings $A_\parallel$>20kHz.
This limitation can be overcome by increasing the total duration of the DDRF sequence. 

\section{Nuclear spin cluster in 4H-SiC}
\begin{figure*}
    \centering
    \includegraphics[width=\textwidth]{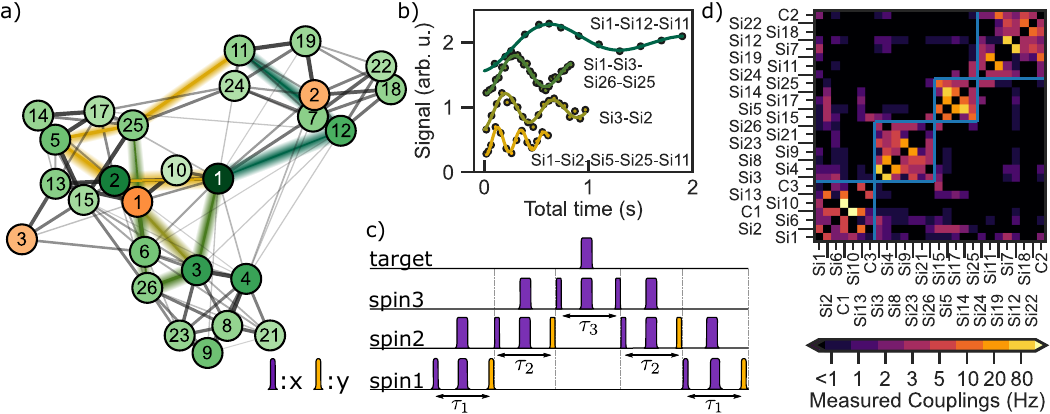}
    \caption{\textbf{a)} Graph representation of the 27-spin network. Each circle represents an isotope (green for silicon, orange for carbon). Grey lines indicate the extracted oscillation frequencies obtained via SEDOR. Couplings below 1 Hz are omitted for clarity. The highlighted links denote the spin chains analyzed in panel b. \textbf{b)} Spin chains. \textbf{c)} Pulse train diagram of nested correlation measurements. \textbf{d)} Experimentally observed oscillation frequencies $f_{ij}$. The spins are arranged such that the emerging clusters (highlighted by blue squares) are grouped. The couplings within each square are used to reconstruct the internal structure of the corresponding cluster, while the couplings outside the squares provide information about the relative spatial arrangement of the clusters with respect to one another.}
    \label{fig:Graph}
\end{figure*}
Understanding the nuclear spin environment is crucial for identifying the primary sources of electron decoherence in solid-state systems.
A well-characterized nuclear spin register enables advanced protocols, such as Loschmidt echo experiments, which can detect and quantify correlated noise and multiparticle entanglement  \cite{brady2024correlated, Macri:2016aa}.
Furthermore, nuclear spins can serve as resources for quantum error correction during entanglement operations \cite{Taminiau_2014, abobeih2022fault}, and pairs of nuclear spins can be manipulated to form decoherence-free subspaces \cite{reiserer2016robust}, thereby enhancing the memory capabilities of the system.
For the NV-center in diamond, mapping a nuclear spin register has facilitated more accurate density functional theory (DFT) calculations, which are often limited by uncertainties—particularly for strongly coupled spins due to the Fermi contact interaction.
High-quality DFT data are essential for high-throughput characterization of spin baths using dynamical decoupling (DD) techniques \cite{poteshman2025trans,varona2025computationally}.

In these systems, nuclear spins are randomly distributed across the crystal lattice sites.
When the dipolar interaction between two nuclear spins is sufficiently strong, it provides a quantitative measure of their spatial separation.
The coupling strength between the two nuclear spins in an unpolarized spin bath can be extracted using spin-echo double-resonance (SEDOR) spectroscopy (Fig. \ref{fig:sedor_spectrum}b), a technique that will be discussed in more detail below.

After initializing the probe spin ($p$), a Hahn-echo sequence is applied, and the accumulated phase is subsequently measured.
The target spin ($t$) is re-coupled by applying a spin-flip operation simultaneously with the echo pulse on $p$, while environmental fluctuations are efficiently refocused.
This procedure selectively interrogates the dipolar interaction between the two spins.
Repeating the sequence while sweeping the frequency of the target-spin pulse yields a spectrum of all nuclei that are dipolar-coupled to the probe spin at a fixed inter-pulse delay $\tau$ (Fig. \ref{fig:sedor_spectrum}a).
Alternatively, varying $\tau$ at a fixed target-spin frequency produces an oscillatory modulation, $f_{ij}$, whose frequency reflects the dipolar coupling strength between the spins (Fig. \ref{fig:sedor_spectrum}d).

\begin{figure*}
    \centering
    \includegraphics[width=\linewidth]{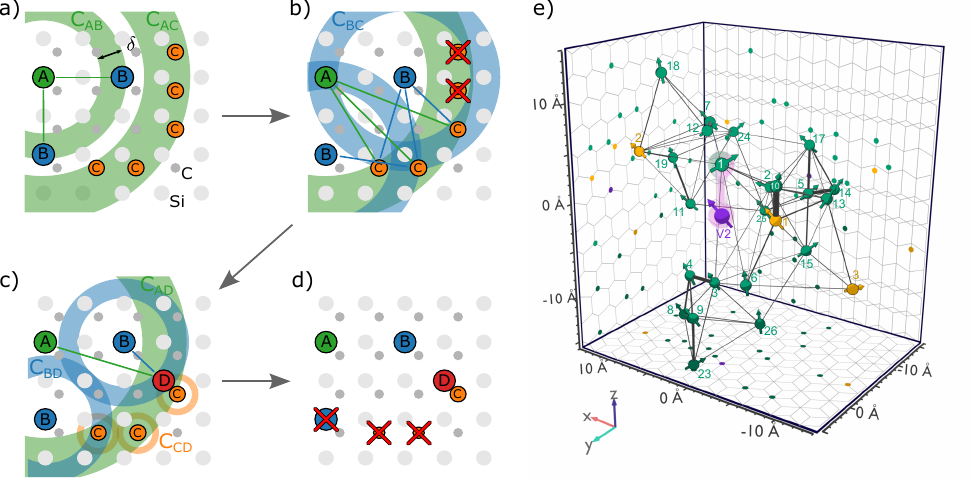}
    \caption{\textbf{a)} Starting from spin A, multiple position candidates for spin B and C are marked based on $C_{AB}$ and $C_{AC}$. \textbf{b)} Including $C_{BC}$ eliminates two marked positions of spin C. \textbf{c)} Spin D shows strong coupling to C. This allows only single positions of D relative to C. \textbf{d)} Based on $C_{AD}$ and $C_{BD}$ two positions of D can be ruled out. \textbf{e)} Spin positions of 25 nuclear spins around central electron spin found by an iterative placement algorithm. Couplings below 2 Hz are omitted for clarity}
    \label{fig:localization}
\end{figure*}

For the majority of the detected nuclear spins, the relevant transition frequencies are well isolated from those of other nuclei.
In those cases, individual spins can be selectively addressed with negligible crosstalk, and SEDOR measurements yield clear oscillations whose frequencies directly correspond to the pairwise dipolar couplings.
This situation allows for straightforward and unambiguous identification of spin pairs.

When probing isotopes that are only weakly coupled to the central electron spin, the resulting SEDOR spectra become increasingly congested.
Multiple nuclei located at distinct lattice sites may possess nearly identical transition frequencies, complicating their unambiguous identification.
Nevertheless, reliable assignment of a specific probe–target pair remains possible when the two spins are in close proximity.
In this regime, the probe spin can be selectively addressed by precisely tuning the interaction time $\tau$.
Bath spins that are located further away—while sharing the same transition frequency as the target spin—couple only weakly to the probe and therefore contribute primarily to a faster decay of the signal.
Probe–target pairs with mutual interaction stronger than this decay rate can still be clearly distinguished and characterized.

As discussed above, in the present system spectral crowding poses significant challenges for selectively addressing individual nuclear spins using DDRF techniques.
This limitation reduces the number of nuclei that can be directly employed as probe spins.
To overcome this constraint and expand the accessible set of probes, we implement a spin-chain protocol (Fig.  \ref{fig:Graph}c). This protocol is based on nested SEDOR blocks, following the approach described in \cite{Biteri_Uribarren_2023, van2024mapping}.

By chaining these correlation steps together, information can be propagated along the nuclear-spin network.
As a result, spins that initially cannot be directly addressed, either because they lie in a spectrally crowded region or due to their direct coupling to initial probe spin is too weak, can still serve as intermediate probes.
This enables the characterization of inter-cluster couplings that would otherwise remain inaccessible (Fig. \ref{fig:Graph}a,b).
All experimentally extracted couplings obtained using this method are summarized in Fig. \ref{fig:Graph}d and used for full 3-dimensional reconstruction of the spin register.

\subsubsection{Three-dimensional nuclear spin localization}
A direct least-squares minimization of the residuals between measured $f_{ij}^{exp}$ and theoretical $f_{ij}^{th}(r_i,r_j)$ oscillation frequencies, computed from dipolar couplings and lattice positions, is not feasible due to the large parameter space and numerous local minima.
Therefore, to reconstruct the three-dimensional positions of the nuclei, we employ an iterative placement algorithm that generates an initial configuration \cite{abobeih2019atomic}: we compare the measured couplings with the dipolar interactions expected from the known crystal lattice geometry.
Starting from the on-axis Si1, we place candidate nuclear spins one by one on the crystallographic lattice (Fig. \ref{fig:localization}a-d). 
At each step, we constrain new spins to sites that yield dipolar couplings to already placed nuclei that match experimental values within a threshold $\delta$ (see Supplementary Material).
In most cases, multiple lattice sites match the dipolar coupling. 
For each new possible position we branch a new set of spin locations. 
After an initial increase of possible spin arrangements, the total number decreases again, because every additional spin restrains the solution further.
However, when a spin from another sub-cluster of spins is added, number of solutions locally grows, until the cluster is increased in size to be rigidly defined.
We increase the number of spins in the solution until only a single spin arrangement is found, which is used as an initial guess for a least-squares minimization. In this final optimization step, the nuclear coordinates are treated as free parameters without lattice discretization.
On average, the final coordinates are shifted by $0.6\,$\AA{}, while the largest shift (Si18) is still below the nearest-neighbour Si-Si distance  distance of $\sim 3.08$\AA{}, indicating atomic precision of the found solution.

The combination of SEDOR spectroscopy with iterative three-dimensional reconstruction enables precise nuclear spin mapping even in the presence of significant spectral crowding.
The reconstruction process reveals a hierarchical organization of the nuclear spin network.
Strongly coupled nuclei ($A_{zz}\gtrapprox10\,$kHz) form a well-resolved basis that defines the core subclusters, providing a robust reference framework for the placement of additional spins.
In contrast, more weakly coupled or spectrally overlapping nuclei present increasing challenges for the localization.
First, the reduced coherence time $T_2$ of weakly coupled nuclei restricts the number of experimentally  resolvable dipolar couplings to $f_{ij}>T_2^{-1} \approx 3$ Hz.
The shortened coherence time arises primarily from an enhanced rate of nuclear spin flip–flop processes, which increases with the number of energetically near-degenerate nuclear spins.
Furthermore, in spectrally crowded regions the finite bandwidth of radio-frequency pulses leads to unintended excitation and thus recoupling of neighboring resonances, further reducing the effective coherence and reducing the observable contrast.
Secondly, to resolve spectrally overlapping spins (e.g. Si8 - Si18 and Si21- Si22), knowledge about the involved subclusters is necessary.
As an illustrative example, after Si9 and Si12 are located, a spin denoted as Si8' must be placed in the lattice.
For the measured couplings $f_{\mathrm{Si8'-Si9}} = 4.31$\,Hz and $f_{\mathrm{Si8'-Si12}} = 4.87$\,Hz no single lattice site can simultaneously reproduce both interactions, because Si9 and Si12 are separated by a distance that precludes a common coupling partner with these strengths.
Thus, we conclude that Si8' consists of two separate spin Si8 and Si18.
This assignment is independently verified by spin-chain measurements that selectively address either Si8 or Si18 and probe their respective coupling environments.

\section{Discussion}
The 25 nuclear spin cluster characterized in the current work contains twenty two $^{29}$Si and three $^{13}$C spins. 
It is consisting out of 4 distinct sub-clusters, each having 5-7 spins. 
We encounter three strongly inter-coupled spin pairs, and one spin-triple, namely Si3-Si4, Si7-Si12, C1-Si10, and Si17-Si5, Si5-Si14, with couplings \{80.06, 80.80, 185.61, 35.80 and 79.65\}\,Hz, respectively. 
Apart from previously reported monoisotope spin pairs of $^{13}$C \cite{bartling2022entanglement}, our system now hosts two species spin pairs, making it an attractive platform for exploring decoherence free subspaces with universal coherent control. 
The cluster like fashion of the register enabled its better exploration, as well as it forms a natural layout for connectivity in the register for potential applications in quantum simulation of spin Hamiltonians. 

The localized register can be further utilized as a backbone for a tree-like search using the spin-chain approach for extension of the register size. 
In the current work we mostly used stronger coupled nuclei ($A_{zz} \gtrapprox 10$\,kHz) to explore deeper into the cluster via spin chaining. 
Due to the topology of those probe spins, in our SEDOR correlation spectroscopy we resulted having all the spins located in a semisphere $y < 0.2$ nm, leaving an empty space of potentially unexplored spin locations and additional potential for extending the register.  
In the next steps, the full initialisation, readout and entangling operations on the register shall be performed. 
In particular, measurement based schemes for two qubit entanglement as well as error correction protocols rely on the highly efficient and fast projective readout. 
In the current state, pure resonant optical readout has a low success rate due to fluorescence intensity, and requires improvement of the photonic collection efficiency of the system via e.g. integration with nanophotonics. 
A reduced application of $N = 10 - 100$ repetitive readouts, is sufficient to bring the efficiency of the readout close to 50\,\%, without degrading the heralding fidelity, which should be explored in the future work.

\section{Conclusion}
In summary, using the achieved highly efficient readout with fidelity exceeding 99 \% we performed extensive NMR spectroscopy of nuclear spin cluster surrounding single V2 color center. 
We measure the hyperfine and internuclear spin coupling parameters of the cluster.
Using them we could localize the spatial positions of the individual nuclear spin within the crystal lattice with single site accuracy. 
We benchmark the hyperfine parameters of the localized spins against the density functional theory predictions and find them quantitatively matching first principle calculations. 
The remaining discrepancy could be utilized as an ethalon for DFT model further relaxation \cite{takacs2024accurate}. 
Together this results unlock the full potential of the color center in silicon carbide platform for applications with local nuclear spin register operation, making a way for future experiments, using a faster, data driven localization algorithms \cite{poteshman2025trans}, validating the numerical predictions of decoherence models \cite{PhysRevB.108.075306}, and benchmarking of novel quantum algorithms tailored for central spin layout \cite{finsterhoelzl2022benchmarking}. 
\section{Acknowledgments}
P.K., J.U.H., and J.W. acknowledge support from the European Commission through the QuantERA project InQuRe (Grant agreements No. 731473, and 101017733). T.S., J.U.H., V.V., and J.W. acknowledge support from the European Union’s Horizon Europe research and innovation program through the SPINUS project (Grant agreement No. 101135699). P.K. and J.W. acknowledge the German ministry of education and research for the project InQuRe (BMBF, Grant agreement No. 16KIS1639K). J.W. acknowledge support from the European Commission for the Quantum Technology Flagship project QIA (Grant agreements No. 101080128, and 101102140), the German ministry of education and research for the project QR.X (BMBF, Grant agreement No. 16KISQ013) and Baden-Württemberg Stiftung for the project SPOC (Grant agreement No. QT-6). J.W. also acknowledges support from the project Spinning
(BMBF, Grant agreement No. 13N16219) and the German Research Foundation (DFG, Grant agreement No. GRK2642). J.U.H. further acknowledges support from the Swedish Research Council under VR Grant No. 202005444 and Knut and Alice Wallenberg Foundation (Grant No. KAW 2018.0071).

\bibliographystyle{apsrev4-2}
\bibliography{bibliography}

	

\end{document}



\title{Supplementary Materials for Localization and coherent control of 25 nuclear spins in Silicon Carbide}
\author{Pierre Kuna}
\thanks{These two authors contributed equally}
 \affiliation{3rd Institute of Physics, IQST, and Research Center SCoPE, University of Stuttgart, Stuttgart, Germany}
\author{Erik Hesselmeier-Hüttmann}
\thanks{These two authors contributed equally}
 \affiliation{3rd Institute of Physics, IQST, and Research Center SCoPE, University of Stuttgart, Stuttgart, Germany}
\author{Phillip Schillinger}
 \affiliation{3rd Institute of Physics, IQST, and Research Center SCoPE, University of Stuttgart, Stuttgart, Germany}
\author{Felix Gloistein}
 \affiliation{3rd Institute of Physics, IQST, and Research Center SCoPE, University of Stuttgart, Stuttgart, Germany}
 \author{István Takács}
 \affiliation{Eötvös Loránd University, Egyetem tér 1-3, H-1053 Budapest, Hungary}
\affiliation{MTA–ELTE Lendület “Momentum” NewQubit Research Group, Pázmány Péter, Sétány 1/A, 1117 Budapest, Hungary}
 \author{Viktor Ivády}
 \affiliation{Eötvös Loránd University, Egyetem tér 1-3, H-1053 Budapest, Hungary}
\affiliation{MTA–ELTE Lendület “Momentum” NewQubit Research Group, Pázmány Péter, Sétány 1/A, 1117 Budapest, Hungary}
 \author{Wolfgang Knolle}
 \affiliation{Department of Sensoric Surfaces and Functional Interfaces, Leibniz-Institute of Surface Engineering (IOM), Permoserstraße 15, 04318 Leipzig, Germany}
\author{Jawad Ul-Hassan}
 \affiliation{Department of Physics, Chemistry and Biology, Linköping University, Olaus Magnus väg, 583 30 Linköping, Sweden}
\author{Jörg Wrachtrup}
 \affiliation{3rd Institute of Physics, IQST, and Research Center SCoPE, University of Stuttgart, Stuttgart, Germany}
 \affiliation{Max Planck Institute for solid state physics, Heisenbergstraße 1, 70569 Stuttgart, Germany}
\author{Vadim Vorobyov}
\email[Correspondence:]{v.vorobyov@pi3.uni-stuttgart.de}
 \affiliation{3rd Institute of Physics, IQST, and Research Center SCoPE, University of Stuttgart, Stuttgart, Germany}

\maketitle
\onecolumngrid
\section{Magnetic field alignment}
\begin{figure}[b]
    \centering{\includegraphics[width=0.5\textwidth]{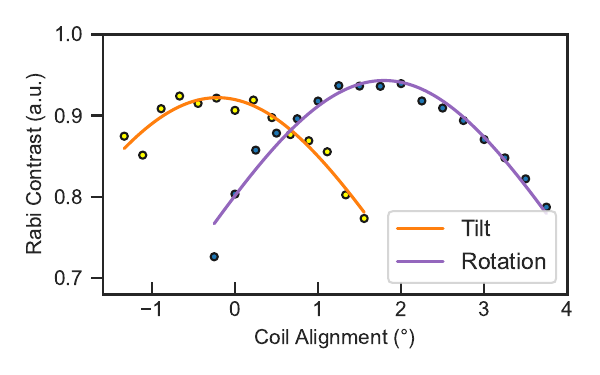}}
    \caption{Magnetic field alignment is performed by measuring the contrast of nuclear spin rabi oscillations, while rotating and tilting the magnetic coils.}
    \label{fig:alignment}
\end{figure}

The applied magnetic field has a magnitude of 1960.9\,G.
Alignment of the field was achieved by rotating the external electromagnet, which consists of two coils with the cryostation positioned between them.
Because the contrast of the SSR scheme is highly sensitive to the external field orientation, the readout contrast was recorded for a range of alignment angles.
Fig. \ref{fig:alignment} shows the SSR contrast as a function of the magnetic-field angle.
A Lorentzian fit to the data yields a $1\sigma$ deviation of $0.037^\circ$ for rotational misalignment and $0.056^\circ$ for tilt misalignment.
These values correspond to a transverse magnetic field of 
\begin{align}
    B_{\perp,\text{rot}} & = \sin(0.037\cdot\pi/180)\cdot1960.9\,\text{G}\approx1.3\,\text{G}\\
    B_{\perp,\text{tilt}} & = \sin(0.056\cdot\pi/180)\cdot1960.9\,\text{G}\approx1.9\,\text{G}\\
    B_\perp &= \sqrt{B_{\perp,\text{rot}}^2+B_{\perp,\text{tilt}}^2} \approx 2.3\,\text{G}
\end{align}

\newpage

\section{Spin relaxation time of on-axis nuclear spin Si1}
\begin{figure}[b]
    \centering{\includegraphics[width=0.95\textwidth]{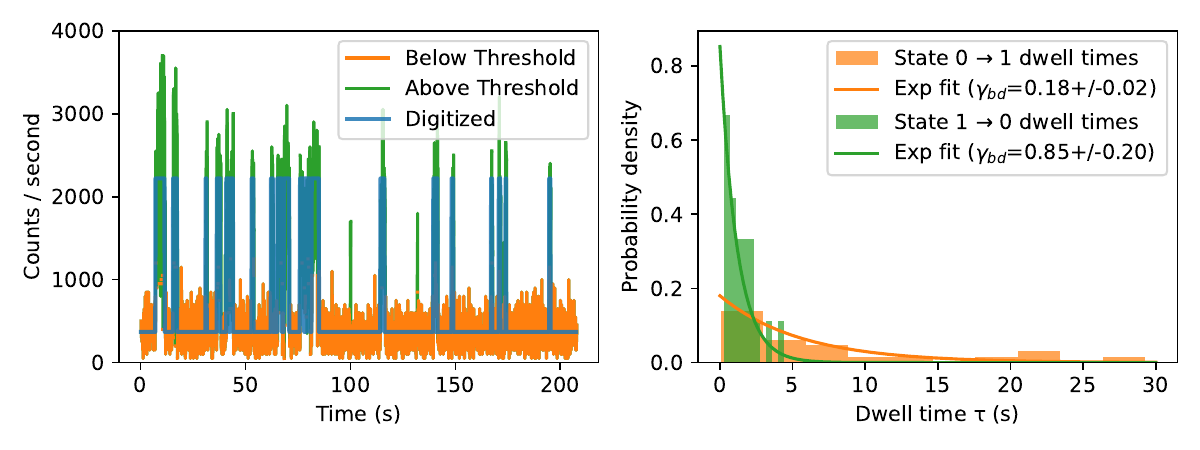}}
    \caption{Left: Time trace of continuous repetitive readout. Right: Time spend in bright and dark state, respectively. Fitting of the dwell times by an exponential function reveals the system's flipping rates.}
    \label{fig:dwelltimes}
\end{figure}

In previous work \cite{Hesselmeier2024_SSR}, the nuclear-spin flipping rates under repeated electron-spin excitation were extracted by fitting the photon-count histogram with an appropriate stochastic model.
This model incorporates the photon count rates of the bright and dark states, as well as the corresponding transition rates between them \cite{zahedian2023readout}.
For the on-axis $^{29}$Si nuclear spin investigated here, the $T_1$ time under readout exceeds the total duration required for N repetitions of the SSR readout sequence.
Consequently, the flipping dynamics cannot be reliably inferred from photon-count histograms alone.
To obtain both transition rates, we therefore record a continuous time trace of the photon signal over \SI{200}{\second} while running the SSR protocol.
The signal is smoothed using a running average to avoid artificially shortened dwell times due to photon-shot-noise fluctuations.
The trace is then discretized by assigning values above (below) 1295 cts/s to the bright (dark) state.
To extract the flipping rates from this long telegraph-like signal, we determine the dwell times spent in each state, as shown in the right panel of Fig. \ref{fig:dwelltimes}.
Assuming a Markovian environment, the dwell-time distribution in each state follows an exponential decay characterized by a single transition rate.
Exponential fits to the dwell-time histograms therefore yield the corresponding bright $\leftrightarrow$ dark flipping rates.
We obtain the following flipping rates:
\begin{align*}
    \text{bright}\rightarrow\text{dark}&: \gamma_{bd} = 0.18\pm0.02\text{Hz}\\
    \text{dark}\rightarrow\text{bright}&: \gamma_{db} = 0.85\pm0.21\text{Hz}
\end{align*}
Thus, we conclude that even under continuous repetitive readout, the nuclear-spin relaxation times are on the order of $\sim$\SI{1}{\second} in the bright state and $\sim \SI{5}{\second}$ in the dark state.

\newpage

\subsection{ENDOR measurements}
Initial characterization of the nuclear spins was carried out using ENDOR (Electron–Nuclear DOuble Resonance) measurements. The experimental procedure is described in detail in Ref.~\cite{Hesselmeier2024_SSR}. During the free evolution between each set of CNOT gates, a phase $\propto A_\parallel$ is written onto the $e^-$-Si1 entangled superposition state.
From these measurements, the transition frequencies of Si2 through Si7, Si9, Si12, and C1 were identified. These more strongly coupled nuclei are located relatively close to the central electron spin and therefore provide the foundation for the subsequent nuclear-spin localization.

The ENDOR spectra were recorded in the $m_s={-1/2, +1/2}$ manifold. In this configuration, nuclear spins with weaker hyperfine coupling ($m_sA_\parallel < 1/T_{2,e}$) cannot imprint a distinct phase onto the electron, which constrains the achievable spectral resolution. When probing the $m_s={-3/2, +3/2}$ manifold, the electron–nuclear interaction strength increases by a factor of three, thereby enabling the detection of a larger number of nuclear spins.
Still, to fully overcome the resolution limit imposed by the electron coherence time, DDRF and SEDOR (Spin-Echo DOuble Resonance) measurements were performed as described in the main text.
\begin{figure}[b]
    \centering{\includegraphics[width=0.95\textwidth]{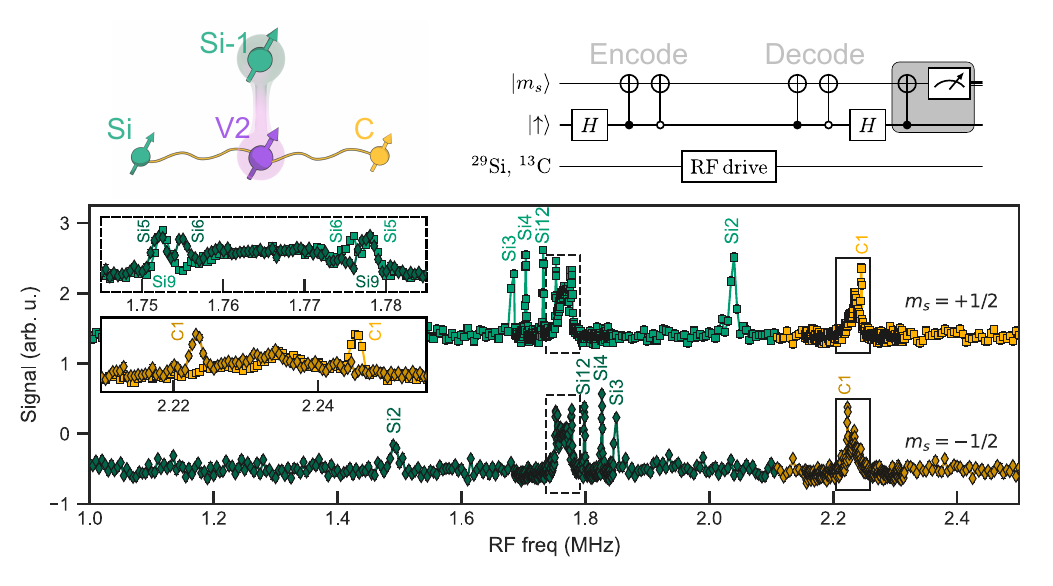}}
    \caption{ENDOR measurements in $m_s={-1/2, +1/2}$. Insets are zoom into the regions around silicon and carbon bath. Due to too weak $A_\parallel$, no isolated signal can be obtained, limiting the usage of this protocoll to spins with $A_\parallel > T_{2,e}$}
    \label{fig:ENDOR}
\end{figure}
\newpage
\section{DDRF sequence}
Dynamical decoupling sequences are typically used to protect the spin qubits from external noise. 
During the nuclear spin control often a significant unwanted noise is imposed onto an electron spin. 
Thus, during the RF gates applied on nuclear spin, an additional dynamical decoupling is applied to the central electron spin. 
Fine tuning of the control field parameter and ability to address nuclear spins conditional to the electron spin subspace by frequency enable creation of unitary gates using such an approach \cite{Bradley2019, Beukers2025, vanOmmen2025}.
To this end, $N$ microwave pulses are applied with sequence as $(\tau-\pi-2\tau-\pi-\tau)^{N/2}$ with resonant radio frequency driving during the interpulse delays of $2\tau$.
The behavior of the system is derived using the Hamiltonian in the interaction picture:
\begin{equation}
\hat{H} = \gamma_n B \cdot I_z + S_z A_{zz} I_z + S_z A_{zx} I_x + 2 \Omega \cos (w_{\mathrm{rf}}t + \phi_{\mathrm{rf}}) I_x,
\end{equation}
where $S_z = -1_e + \sigma_z/2$ is the pseudospin matrix in the $m_s = -3/2, -1/2$ subspace denoted as $|0\rangle, |1\rangle$, and $I_x, I_z$ are the spin matrices of nuclear spin $\hat{I} = 1/2$. 
For simplicity we neglect the $A_{zx}$ term, and obtain the simplified form: 
\begin{equation}
\hat{H} = \omega_0 |0\rangle \langle 0| \otimes I_z + \omega_1 |0\rangle \langle 0| \otimes I_z +  2 \Omega \cos(\omega _{\mathrm{rf}}t + \phi_{\mathrm{rf}}) 1 \otimes I_x,
\end{equation}
where $\omega_0 = (\gamma_n B - 3/2 \cdot A_{zz})$ and $\omega_1 = (\gamma_n B - 1/2 \cdot A_{zz})$ for respective subspaces. 
Using the rotating frame transformation with $U = \exp( i \omega_{\mathrm{rf}} t I_z)$, and expanding $\cos(\omega_{\mathrm{rf}}t + \phi_{\mathrm{rf}}) = e^{iw_{\mathrm{rf}}t} (\cos \phi_{\mathrm{rf}} + i \sin \phi_{\mathrm{rf}})$, the net interaction picture Hamiltonian is thus: 
\begin{equation}
H_I = U H U^\dag - i \frac{dU}{dt}U^\dag = \Delta_0 |0 \rangle\langle 0| \otimes I_z + \Delta_1 |1 \rangle\langle 1| \otimes I_z + \Omega \cdot 1_e \otimes (I_x \cos \phi_{\mathrm{rf}} + I_y \sin \phi_{\mathrm{rf}}),
\end{equation}
where $\Delta_i = \omega_i - \omega_{\mathrm{rf}}$. 
By choosing the driving frequency resonant to e.g. sub-level resonances, $\omega_{0}$, the nuclear spin is conditionally being driven for electron spin state $|0\rangle$ and freely precessing in the $|1\rangle$ subspace, provided that the driving strength $\Omega \ll \Delta_{1}$.  
Furthermore, to achieve the conditional rotation during the periodic flip of the electron spin the nuclear spin drive is engineered via the phase of the radiofrequency $\phi_{\mathrm{rf}}$, which is shifted by $\pi$ and creates opposite direction of rotation for even and odd rf-pulses. 
This results in overal conditional evolution of nuclear spin dynamics on initial state of the electron spin $|0\rangle$ or $|1\rangle$. 
During the idling periods, the precession of the nuclear spin has to be accounted, $\phi_{acc} = (\omega_0 + \omega_1 - 2 \omega_{\mathrm{rf}}) \cdot 2\tau$, due to off-resonant evolution. 
Thus, the general phase update expression can be obtained: 
\begin{equation}
\delta_{\phi, \mathrm{rf}} = \pi + (\omega_0 + \omega_1 - 2\omega_{\mathrm{rf}}) \tau
\end{equation}
Furthermore, due to off-resonant driving effects of pulses of length $\tau$, the apparent pulse side peaks in Fourier-space $\sim 1/\tau \gg \Omega$ are larger than power broadening \cite{vanOmmen2025}, the effective Rabi drive frequency reads for the case of rectangular pulses $\Omega_{eff} = \Omega \left[ \mathrm{sinc}\left((\omega_1 - \omega_{\mathrm{rf}}) \tau \right) -\mathrm{sinc}\left((\omega_0 - \omega_{\mathrm{rf}}) \tau \right)\right]$, which yields the overall rotation angle $\theta_N = \pm N\Omega_{eff}\tau$. Sign of the rotation is conditioned on the $\textit{initial}$ electron state at the start of the sequence. 
The final phase of the nuclear spin state depends on the precession during the whole sequence $\phi = N\phi_{acc}(\omega_{\mathrm{rf}}, \tau)$.
When chosen $\theta_N = \pm \pi/2$ a maximally entangling gate, resulting in a CNOT gate up to a global phase of the nuclear spin. 
By adjusting the $\tau$ inter-pulse timing with respect to the effective precession frequency, $\tau = n \pi /\omega_{eff}$ the global phase can be cancelled. 
Furthermore, by adjusting the detuning, phase increment and driving strength an optimized gate accounting for $A_{zx,zy}$ could be realized and is a subject for further work  \cite{regina_ddrf}.

\section{Spin placement}
The three-dimensional reconstruction of the nuclear-spin cluster relies on measurements of the internuclear interactions $I^{(i)}CI^{(j)}$.
The dominant contribution to this interaction is the $z$-component of the dipolar coupling, $C_{zz}$, which for two nuclear spins $i$ and $j$ is given by
\begin{equation}
    C_{ij} = \frac{\alpha_{ij}}{\Delta r_{ij}^3}\left(\frac{3(z_j-z_i)^2}{\Delta r_{ij}^2}-1\right),
    \label{eq:dipolar}
\end{equation}
where $\Delta r_{ij} = \sqrt{(x_j-x_i)^2+(y_j-y_i)^2+(z_j-z_i)^2}$, $\alpha_{ij} = \mu_0\gamma_i\gamma_j\hbar/4\pi$, $\mu_0$ is the vacuum permeability, $\gamma_i$ is the gyromagnetic ratio of spin $i$ and $\hbar$ is the reduced Planck constant.
Because $C_{ij}$ depends exclusively on the relative position vector between the two spins, knowledge of $C_{ij}$ constrains their spatial separation and orientation.
In a SEDOR experiment, the oscillation frequency is related to the magnitude of the coupling by $f_{ij} \approx |C_{ij}|/2$.
By acquiring a sufficiently large set of internuclear couplings, the reconstruction problem becomes overdetermined, enabling precise determination of the nuclear-spin coordinates.
During the placement procedure each nuclear spin is restricted to its corresponding sublattice (Si or C).
Consequently, the relevant nearest-neighbor spacing for allowed nuclear locations is not the Si–C bond length ($\approx 1.91$\AA), but the distance between equivalent lattice sites, namely the nearest Si–Si or C–C separation ($\approx 3.08\AA$). This drastically reduces the amount of possible lattice sites per coupling compared to spin localization within a diamond lattice with bond length of $\approx 1.54$\AA.

For embedding the spin cluster into the 4H-SiC lattice, we chose the position of V2 center as the origin of the coordinate system $(x,y,z) = (0,0,0)$.
Comparison with density-functional theory (DFT \cite{Hesslemeier2024_GSLAC}) uniquely identifies the lattice position of the on-axis Si1 at $(x,y,z) = (0,0,c/2)$, where $c$ is the lattice constant along the crystallographic $c$-axis. This serves as the initial reference point for the iterative spin-placement procedure.
For the placement of the next nuclear spin $j$ the corresponding internuclear coupling $C_{ij}$ is extracted from the SEDOR measurements. To incorporate experimental uncertainties as well as deviations arising from perturbative corrections, we apply a tolerance window of typically $\delta=\pm0.6$\,Hz (see detailed discussion below). Using this tolerance, all lattice sites whose dipolar coupling to an already placed reference spin $i$ is consistent with the measured $C_{ij}$ are considered as candidate positions for spin $j$. Those candidates are then crosschecked with potential other already placed spins, further reducing the count of canditates.
Each valid spin arrangement generated at this stage is retained as an independent partial solution, which subsequently serves as the basis for placing the next spin in the sequence.

\begin{figure}
    \centering{\includegraphics[width=0.95\textwidth]{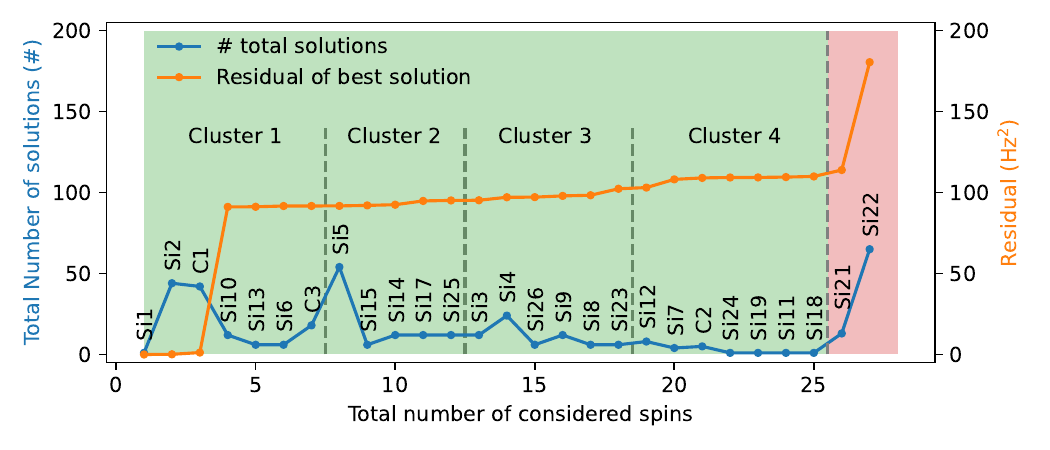}}
    \caption{Blue: Total number of spin-placement solutions after iteratively constraining  the position of $N$ spins. Orange: Remaining residual $\epsilon = \sum_{i,j}  (f_{ij,\mathrm{exp}} - f_{ij,\mathrm{th}})^2$ for the best found solution at each $N$ placed spins. With the placement of Si10 the residual drastically increases. Still, due to the very strong coupling of $f_{ij} = 186.3$\,Hz to C1 it can be uniquely located. Green area marks spins which could be located unambiguously after placing 25 spins, red area marks spins which show multiple solutions. Si21 has 13 different possible lattice sites. Si22 has 6 different possible lattice sites.}
    \label{fig:numberSolutions}
\end{figure}

\begin{figure}
    \centering{\includegraphics[width=\textwidth]{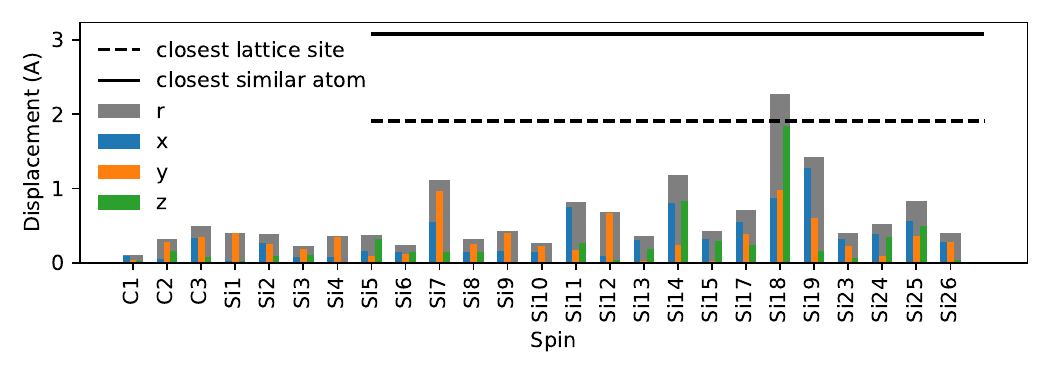}}
    \caption{$x,y,z$- and total displacement of each nuclear spin after the least squares minimization. An average deviation of 0.6\,\AA{} is obtained. The largest displacement of 2.3\,\AA{} (Si18) is still below the distance of two closest silicons ($\approx 3.08$\,\AA{}).}
    \label{fig:leastSquare}
\end{figure}

Initially, the number of possible configurations increases with the number of placed spins. However, as additional spins are incorporated, the set of feasible solutions begins to contract.
This reduction is particularly pronounced when multiple spins occupy the same spatial region of the lattice and form a local sub-cluster.
Adding further spins within such a sub-cluster typically leads to a rapid elimination of incompatible configurations, whereas introducing a spin that lies outside a known sub-cluster generally increases the number of admissible solutions.
After iteratively placing 25 nuclear spins within the 4H-SiC lattice, the algorithm converges to only a single distinct solutions.
Fig. \ref{fig:numberSolutions} illustrates the evolution of the total number of different solutions as a function of the number of spins placed.
Following an initial growth phase, only a single configurations remains once 25 spins have been positioned.
The spins Si21 and Si22 cannot be uniquely assigned and are therefore excluded from the set of fully located spins.
The ordering of the spin placement is chosen such that spatial sub-clusters are resolved first, thereby minimizing the total number of admissible configurations at each iteration. 
Because of the long bond length and the small tolerance of 0.6 Hz, the admissible lattice sites are highly constrained. Consequently, only a limited number of candidate positions must be propagated in each iteration of the algorithm, enabling rapid spin placement within a few seconds on a standard desktop PC.
The resulting 25-spin configuration serves as the initial guess for a subsequent least-squares refinement, in which the nuclear coordinates are no longer constrained to lie on the discrete 4H-SiC lattice sites.
In this continuous relaxation procedure, the spin positions adjust solely according to the measured internuclear couplings. As shown in Fig. \ref{fig:leastSquare}, all spins relax to positions separated by distances smaller than the nearest-neighbor Si–Si or C–C spacing ($\approx 3.08\AA$), indicating atomically precise localization of the reconstructed spin cluster.

\begin{figure}
    \centering{\includegraphics[width=0.95\textwidth]{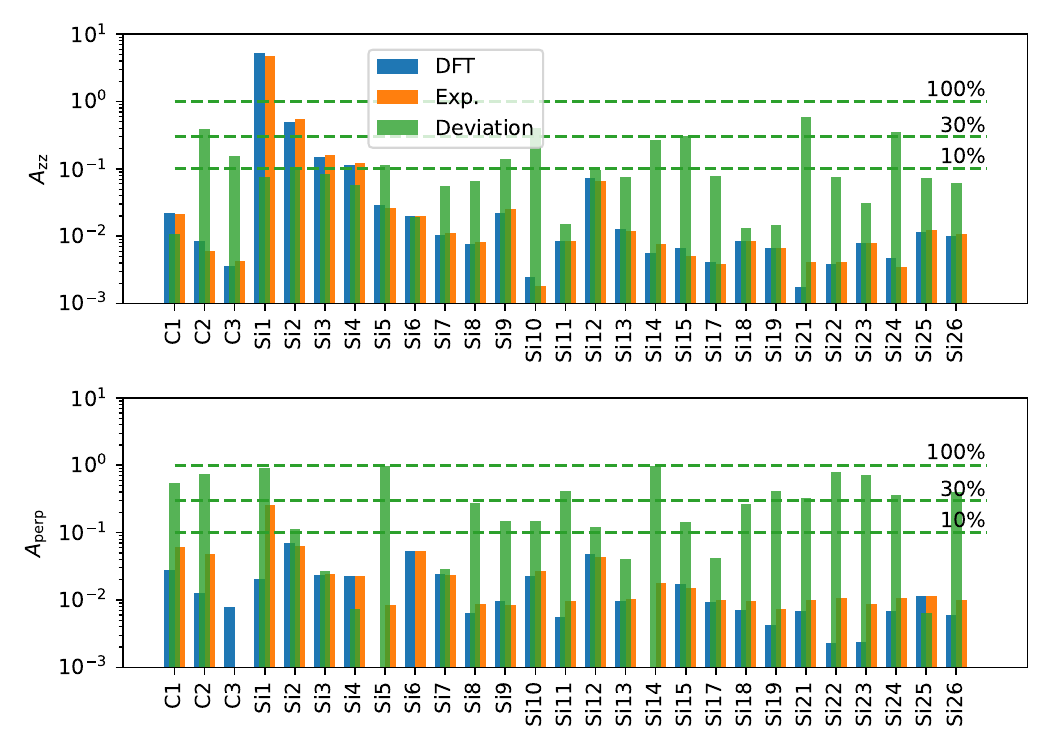}}
    \caption{Comparison of DFT data (blue) and experiment (orange). Green bars indicate the relative deviation. Top: Comparison of $A_{zz}$. Some values show a deviation >30\%, while for most spins the deviation is 10\% or below. For better visibility the absolute value of $A_{zz}$ is shown. The sign of all values agree with each other. Bottom: Comparison of $A_{\perp}$. Here we see slighlty increased relative errors, which is due to weaker sensitivity to $A_{\perp}$ in the experiment. Furthermore, the smaller $A_{\perp}$ becomes, the larger the relative error.}
    \label{fig:comparisonDFT}
\end{figure}

A comparison of the reconstructed spin configuration with density functional theory (DFT) calculations shows excellent overall agreement (Fig. \ref{fig:comparisonDFT}).
For the hyperfine component $A_{zz}$, most spins exhibit deviations of 10\% or less, although a few outliers deviate by more than 30\%.
The transverse hyperfine component $A_\perp$ shows slightly larger relative discrepancies. This behaviour originates from the reduced experimental sensitivity to $A_\perp$.
Furthermore, the smaller $A_{\perp}$ becomes, the larger the relative error.

\begin{figure}
    \centering{\includegraphics{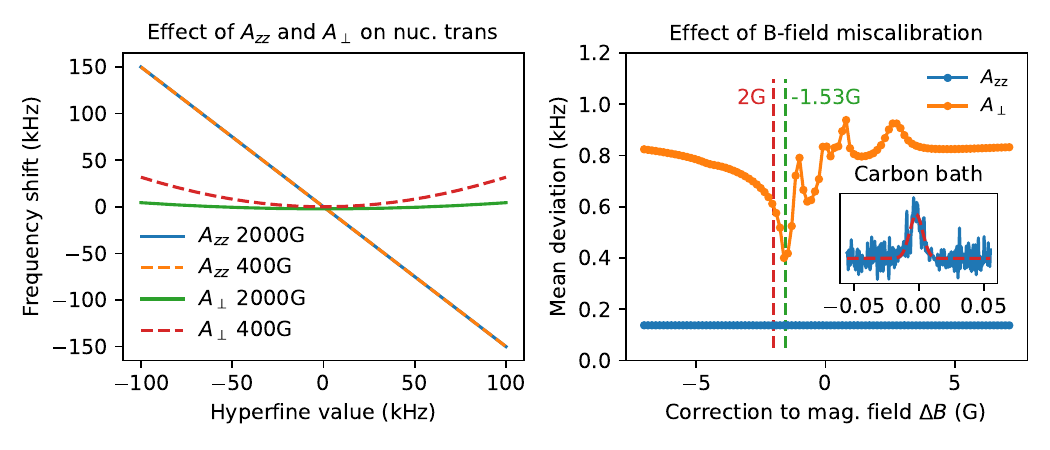}}
    \caption{Left: The sensitivity to $A_\perp$ is reduced for  strong magnetic field. Right: Magnetic field miscalibration leads to incorrect transverse hyperfine parameters, while the longitudinal parameters remain unaffected.}
    \label{fig:Azx_Sensitivity}
\end{figure}

The reduced sensitivity of $A_{\perp}$ originates from the relatively strong external magnetic field of $1960.9$\,G. Nuclear spin transitions frequencies are measured in the electron subdomains $\pm3/2$. By solving 
\begin{equation}
    f_i = \sqrt{(\gamma_i\cdot B + m_s\cdot A_{zz})^2 + (m_s\cdot A_\perp)^2}
\end{equation}
for $A_{zz}$ and $A_{\perp}$ the hyperfine parameters can be extracted.
At high magnetic fields, the term proportional to $A_{zz}$ dominates the expression, suppressing the influence of $A_\perp$ on the measured frequencies.
This reduction in sensitivity is illustrated in the left panel of Fig. \ref{fig:Azx_Sensitivity}.

\subsubsection{Electron g-factor}
We observe a substantial mismatch between measured $A_\perp=\sqrt{A_{zx}^2+A_{zy}^2}$ and values obtained by DFT simulation. In the following we reason that this mismatch could originate from a magnetic field miscalibration. The Landé factor of the central electron spin was assumed to be $g=-2.0028$ during the experiments. The Landé factor depends on defect inherent properties like the spin-orbit coupling and thus local variations like strain can alter this value.

The inset on the right hand side of Fig. \ref{fig:Azx_Sensitivity} shows the carbon bath obtained by SEDOR measurements from Si1.
A Gaussian fit reveals that the center of the bath is shifted by $\Delta f_C =-2.1$\,MHz from $f_C = \gamma_CB$, which translates to $\Delta B_C \approx -2$\,G.
At the present strong magnetic field the distribution of carbon isotope resonances is dominated by their $A_{zz}$ term.
Since the carbon isotopes are randomly distributed in the lattice, it can occur that the present statistical mixture has an increased number of positive or negative $A_{zz}$ couplings, shifting the overall peak.
Thus, a second method to verify a shift of the magnetic field is employed.

Si5 and Si14 are two spins which are located on the same $z$-coordinate as the V2 center.
Thus, their perpendicular coupling should be the weakest of all investigated spins.
Various correction $\Delta B$ are applied and the mismatch between experimental values and DFT data is recalculated. At $\Delta B_{Si5} = -1.47$\,G and $\Delta B_{Si14} = -1.59$\,G the mismatch is minimal with values of $A_\perp = 0.7$\,kHz and $A_\perp = 0.4$\,kHz, for Si5 and Si14 respectively. 
Leading to an average shift of $\Delta_{Si} = -1.53$\,G.

The nuclear transition frequency in the presence of an electron spin manifold $m_s$ can be expanded perturbatively in the hyperfine interaction. To zeroth order one obtains
\begin{equation}
    f^{(0)} = \gamma_nB + m_sA_{zz},
\end{equation}
which shows that the nuclear-spin resonance frequency depends linearly on the longitudinal hyperfine coupling $A_{zz}$.
The first-order correction vanishes,
\begin{equation}
    f^{(1)} = 0,
\end{equation}
because the corresponding matrix elements do not connect states within the nuclear-spin subspace.

The leading contribution of the transverse hyperfine components arises in second order. Neglecting terms proportional to $1/\gamma_eB$, the second-order shift is of the form
\begin{equation}
    f^{(2)} \propto \frac{A_\perp^2}{\gamma_nB + m_sA_{zz}},
\end{equation}
with $A_\perp=\sqrt{A_{zx}^2+A_{zy}^2}$.
This expression reveals a quadratic dependence on the transverse hyperfine interaction and an inverse dependence on the effective nuclear Zeeman splitting. Consequently, the experimentally extracted value of $A_\perp$ is highly sensitive to small variations or calibration uncertainties in the magnetic field $B$.
In contrast, $A_{zz}$ enters only at zeroth order and therefore remains comparatively insensitive to the same variations.
Fig. \ref{fig:Azx_Sensitivity} shows the average deviation of DFT and experimental hyperfine parameters versus the correction to the magnetic field $\Delta B$. A clear minimum is visible at $-1.53$\,G, which is in agreement of the above extracted corrections of $-1.47$\,G and $-1.59$\,G.
With the uncertainty $\Delta B_C-\Delta B_{Si}\approx0.6$\,G a magnetic field correction of $\Delta B = -1.53\pm0.6$\,G at $1960.9$\,G changes the field amplitude by $\approx-0.08(3)\%$.
Thus the electron $g$-factor in the present system can be estimated to be $g=-2.0012(6)$.
\section{Perturbation theory for inter coupling strength}
The system under consideration consists of a central electron spin coupled to two surrounding nuclear spins, each of which may correspond to either the $^{13}$C or $^{29}$Si isotope.
The full spin Hamiltonian of this tripartite system can be decomposed into four contributions.
The electron zero-field splitting and electron Zeeman interaction ($H_e$), the nuclear Zeeman interaction ($H_n$), the electron-nuclear interactions ($H_{en}$) as well as the internuclear coupling ($H_{nn}$):
\begin{align}
H &= H_e + H_n + H_{en} + H_{nn}\\
H_e &= DS_z^2 + \gamma_e(B_xS_x + B_yS_y + B_zS_z)\\
H_n &= \gamma_n^{(1)}B_zI_z^{(1)}+ \gamma_n^{(2)}B_zI_z^{(2)}\\
H_{en} &= S A^{(1)} I^{(1)} + S A^{(2)} I^{(2)}\\
H_{nn} &= I^{(1)} C I^{(2)}\\
\end{align}
A strong magnetic field is applied along the crystallographic $z$-axis, which is chosen as the quantization axis for the perturbative treatment.
Using this axis, the unperturbed Hamiltonian is defined as
\begin{equation}
H_0 = DS_z^2 + \gamma_eB_zS_z + \gamma_n^{(1)}B_zI_z^{(1)} + \gamma_n^{(2)}B_zI_z^{(2)} + S_zA_{zz}^{(1)}I_z^{(1)} + S_zA_{zz}^{(2)}I_z^{(2)} + I_z^{(1)}C_{zz}I_z^{(2)}.
\end{equation}
whose eigenstates are $\psi_i = |m_s, m_I^{(1)}, m_I^{(2)}\rangle$, with $m_s = \{\pm3/2, \pm1/2\}$ and $m_I = \{\pm1/2\}$.
The eigenenergies are given by
\begin{equation}
\lambda_0(m_s, m_I^{(1)}, m_I^{(2)}) = m_s^2D + \gamma_eB_zm_s + \gamma_n^{(1)}B_zm_I^{(1)} + \gamma_n^{(2)}B_zm_I^{(2)} + m_sm_I^{(1)}A_{zz}^{(1)} + m_sm_I^{(2)}A_{zz}^{(2)}+ m_I^{(1)}m_I^{(2)}C_{zz}.
\end{equation}

The oscillation frequency observed in a SEDOR experiment for a fixed electron spin projection $m_s$ is given by
\begin{align}
f_\text{SEDOR}(m_s) &= \frac{1}{2}\left|\lambda\left(m_s, \frac{1}{2}, \frac{1}{2}\right) + \lambda(m_s, -\frac{1}{2}, -\frac{1}{2}) - \lambda(m_s, -\frac{1}{2}, \frac{1}{2}) - \lambda(m_s, \frac{1}{2}, -\frac{1}{2})\right|\\
f_\text{SEDOR}(m_s)& \approx \frac{1}{2}|m_sD + m_s\gamma_eB_z + \frac{1}{2}\gamma_n^{(1)}B_z + \frac{1}{2}\gamma_n^{(2)}B_z + m_s\frac{1}{2}A_{zz}^{(1)} + m_s\frac{1}{2}A_{zz}^{(2)} + \frac{1}{4}C_{zz}\\
&+ m_sD + m_s\gamma_eB_z - \frac{1}{2}\gamma_n^{(1)}B_z - \frac{1}{2}\gamma_n^{(2)}B_z - m_s\frac{1}{2}A_{zz}^{(1)} - m_s\frac{1}{2}A_{zz}^{(2)} + \frac{1}{4}C_{zz}\\
&-m_sD - m_s\gamma_eB_z + \frac{1}{2}\gamma_n^{(1)}B_z - \frac{1}{2}\gamma_n^{(2)}B_z + m_s\frac{1}{2}A_{zz}^{(1)} - m_s\frac{1}{2}A_{zz}^{(2)} + \frac{1}{4}C_{zz}\\
&-m_sD - m_s\gamma_eB_z - \frac{1}{2}\gamma_n^{(1)}B_z + \frac{1}{2}\gamma_n^{(2)}B_z - m_s\frac{1}{2}A_{zz}^{(1)} + m_s\frac{1}{2}A_{zz}^{(2)} + \frac{1}{4}C_{zz}|\\
& = \frac{1}{2}\left|C_{zz}\right|,
\end{align}
for $\lambda = \lambda_0$.
This demonstrates that, at zeroth order, the SEDOR oscillation frequency depends solely on the secular internuclear dipolar coupling $C_{zz}$.
Corrections to this coupling arise from electron-mediated interactions, non-secular dipolar terms, and magnetic-field misalignment.
These contributions are quantified using perturbation theory in the following.
Writing the perturbation as $V = H - H_0$, the first-order correction to the energy of state $\psi_n$ is 
\begin{equation}
\langle\psi_n|V|\psi_n\rangle,
\end{equation}
By construction of $H_0$, all diagonal elements of $V$ vanish, i.e.
\begin{equation}
    \langle\psi_n|V|\psi_n\rangle = 0
\end{equation}
The second-order correction are
\begin{equation}
\lambda(\psi_n)\approx\lambda_0(\psi_n) + \sum_{k\neq n}\frac{\left|\langle\psi_k|V|\psi_n\rangle\right|^2}{\lambda_0(\psi_n)-\lambda_0(\psi_k)}.
\end{equation}
The deviation of the perturbation terms is presented in great detail in the supplementary information of Abobeih et al. \cite{abobeih2019atomic} (equation (S9) -- (S13)). Similar calculations where performed for the present spin-3/2 system in a bi-atomic lattice, leading to 
\begin{align}
    \Delta\lambda_1(m_s = \pm3/2) &\approx \frac{9}{4}\frac{a_{zx}^{(1)}a_{zx}^{(2)}+a_{zy}^{(1)}a_{zy}^{(2)}}{m_s\gamma_e B_z + 2D}\\
    \Delta\lambda_2(m_s = \pm3/2) &\approx m_s\Delta\lambda_2^{(0)} +\Delta\lambda_2^{(1)}\\
    \Delta\lambda_2^{(0)} &= \sum_{j=1}^2\frac{a_{zx}^{(j)}c_{zx}+a_{zy}^{(j)}c_{zy}}{\gamma_jB_z}\\
    \Delta\lambda_2^{(1)} &=-\frac{9}{4}\sum_{j=1}^2a_{zz}^{(j)}\left(\frac{a_{zx}^{(j)}c_{zx}+a_{zy}^{(j)}c_{zy}}{\gamma_j^2B_z^2}\right)\\
    \Delta\lambda_3(m_s = \pm3/2) &\approx \Delta\lambda_3^{(0)} +m_s\Delta\lambda_2^{(1)}\\
   \Delta\lambda_3^{(0)} &= 2\frac{B_xc_{zx}+B_yc_{zy}}{B_z}\\
   \Delta\lambda_3^{(1)} &= \sum_ia_{zz}^{(i)}\frac{(B_xc_{zx}+B_yc_{zy})}{B_z^2\gamma_i}.
\end{align}
This shows that there is no qualitiative difference between the spin-1 system and spin-3/2 system as well as mono- and bi-atomic lattice. Therefore, averaging over the two electron subdomains $\pm3/2$ eliminates $\Delta\lambda_2^{(0)}$ and $\Delta\lambda_3^{(1)}$. Note, that the magnetic field amplitude in the demoniator indicates smaller perturbation contributions the larger the magnetic field is.

\begin{figure}
 \centering{\includegraphics[width=0.95\textwidth]{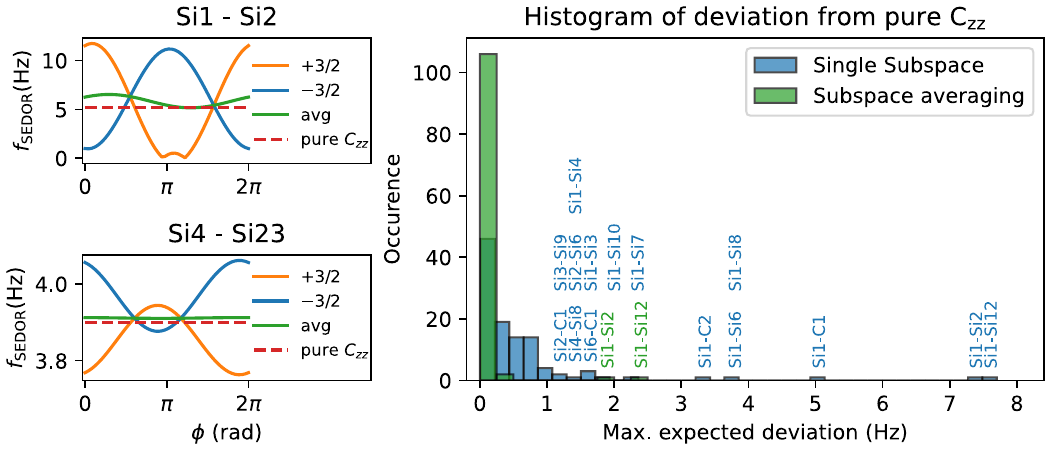}}
 \caption{Maximum deviation between 0$^{th}$-order and 2$^{nd}$-order perturbation theory, where a transverse magnetic field of 2.3\,G is assumed. $\phi$ is used to parameterize $A_{zx} = \cos(\phi)A_\perp$ and $A_{zy} = \sin(\phi)A_\perp$. Averaging over $m_s\{-3/2, +3/2\}$ drastically reduces the perturbation effects.}
 \label{fig:Czz_deviation}
\end{figure}

Figure \ref{fig:Czz_deviation} illustrates the magnitude of perturbative corrections for two representative nuclear-spin pairs: Si1--Si2 and Si4--Si23.
The Si1–Si2 pair exhibits strong deviations from the ideal dipolar coupling due to its strong hyperfine interaction with the central electron, leading to shifts from below <1\,Hz to 10\,Hz.
Averaging over the two electron manifolds $m_s=\pm3/2$ substantially suppresses these deviations, limiting the expected error to approximately 2\,Hz.
In contrast, the second spin pair shows a maximal deviation of only 0.3\,Hz, even without subspace averaging.
The right panel if Fig. \ref{fig:Czz_deviation} summarizes the maximum perturbative deviations for all measured nuclear-spin pairs, both for individual electron subspaces and for the averaged case.
These results justify the heuristic tolerances used in the iterative spin-placement algorithm:
if an internuclear coupling is measured in both electron subspaces, the expected deviation is below 0.6 Hz for all but two pairs (Si1--Si2 and Si1--Si12), for which a larger tolerance of 3\,Hz is employed.
Spin pairs exhibiting large subspace-specific deviations were accordingly analyzed in both subspaces. Furthermore, for large couplings >35\,Hz we use a relative tolerance of ~5\% of the measured coupling.

\begin{table*}
\begin{tabular}{l|cc|cc|cc|rrr}
 Label & $\Delta_{-3/2}$ (kHz) & $\Delta_{+3/2}$ (kHz) & $\omega_{-3/2}$ (kHz) & $\omega_{+3/2}$ (kHz) & $A_\parallel$ (kHz)& $A_\perp$ (kHz) & x (\AA{}) & y (\AA{}) & z (\AA{})\\
\hline\hline
Si1 & 7225(1) & -7233(1) & 8886(1) & -5571(1) & -4811.8(8) & 260(1) & 0.00 & 0.00 & 5.04 \\
Si2 & -814.1(6) & 819(1) & 847.2(6) & 2480(1) & 545.4(4) & 63(3) & -3.08 & -3.56 & 2.52 \\
Si3 & 243.3(2) & -245.1(3) & 1904.6(2) & 1416.2(3) & -162.85(4) & 23(2) & 1.54 & -0.89 & -7.56 \\
Si4 & 183.5(2) & -185.4(3) & 1844.8(2) & 1475.9(3) & -122.99(5) & 21(3) & 4.62 & -0.89 & -7.56 \\
Si5 & -40.06(4) & 37.57(5) & 1621.21(4) & 1698.83(5) & 25.874(8) & 3(2) & -3.08 & -10.67 & 0.00 \\
Si6 & -29.6(2) & 30.9(2) & 1631.7(2) & 1692.1(2) & 20.18(4) & 52.6(8) & -4.62 & 2.67 & -5.04 \\
Si7 & 15.8(2) & -17.6(2) & 1677.1(2) & 1643.6(2) & -11.15(4) & 22(2) & 4.62 & -4.45 & 7.56 \\
Si8 & 11.2(4) & -13.8(4) & 1672.5(4) & 1647.5(4) & -8.35(6) & 9(5) & 6.16 & -1.78 & -12.60 \\
Si9 & 36.7(2) & -39.2(2) & 1698.0(2) & 1622.1(2) & -25.31(4) & 7(4) & 1.54 & 2.67 & -10.08 \\
Si10 & -3.5(1) & 1.9(2) & 1657.8(1) & 1663.2(2) & 1.80(4) & 25(2) & -7.70 & 2.67 & 5.04 \\
Si11 & -14.1(3) & 11.6(5) & 1647.1(3) & 1672.8(5) & 8.56(8) & 10(6) & 9.24 & -7.11 & -2.52 \\
Si12 & 99.9(3) & -100.0(2) & 1761.2(3) & 1561.3(2) & -66.68(5) & 42(1) & 3.03 & -1.84 & 7.43 \\
Si13 & -19.2(2) & 16.8(2) & 1642.0(2) & 1678.1(2) & 12.00(4) & 7(4) & -9.24 & -3.56 & 2.52 \\
Si14 & -13.0(3) & 10(2) & 1648.3(3) & 1672(2) & 7.7(3) & 18(9) & -4.62 & -13.34 & 0.00 \\
Si15 & -8.7(3) & 6.5(5) & 1652.6(3) & 1667.7(5) & 5.06(8) & 12(7) & -4.62 & -8.00 & -5.04 \\
Si17 & -7.0(3) & 4.5(5) & 1654.2(3) & 1665.8(5) & 3.85(8) & 10(5) & -3.08 & -10.67 & 5.04 \\
Si18 & 11.7(4) & -14.2(4) & 1672.9(4) & 1647.1(4) & -8.61(6) & 9(5) & 7.70 & -0.89 & 12.60 \\
Si19 & -11.3(2) & 8.7(2) & 1650.0(2) & 1670.0(2) & 6.67(4) & 8(4) & 10.78 & -6.22 & 2.52 \\
Si21 & 4.9(2) & -7.5(6) & 1666.2(2) & 1653.7(6) & -4.14(10) & 11(7) & NaN & NaN & NaN \\
Si22 & 5.0(4) & -7.6(7) & 1666.3(4) & 1653.6(7) & -4.2(1) & 12(7) & NaN & NaN & NaN \\
Si23 & 10.8(4) & -13.4(4) & 1672.1(4) & 1647.9(4) & -8.07(6) & 10(5) & 1.54 & 2.67 & -15.12 \\
Si24 & -6.5(2) & 4.0(7) & 1654.8(2) & 1665.3(7) & 3.5(1) & 11(5) & 6.16 & -10.67 & 5.04 \\
Si25 & -20.0(3) & 17.6(3) & 1641.3(3) & 1678.8(3) & 12.52(6) & 9(5) & 1.54 & -9.78 & -2.52 \\
Si26 & 15.1(2) & -17.7(6) & 1676.4(2) & 1643.6(6) & -10.93(9) & 11(6) & -1.54 & -4.45 & -12.60 \\
C1 & 32.1(5) & -33.0(5) & -2069.3(5) & -2134.4(5) & -21.73(9) & 61(2) & -7.70 & 2.67 & 1.91 \\
C2 & 9.5(2) & -8.7(6) & -2092.0(2) & -2110.2(6) & -6.07(9) & 47(3) & 10.78 & -0.89 & 4.39 \\
C3 & 6.3(7) & -6(1) & -2095.2(7) & -2107(1) & -3.9(2) & 47(5) & -12.32 & -3.56 & -5.69 \\
\bottomrule
\end{tabular}
\caption{Compact summary of Si and C spin values: Detunings from their respective bath frequencies ($\Delta$), transition frequencies ($\omega$), hyperfine parameters $A_{\parallel}$ and $A_{\perp}$, as well as the reconstruced 3D coordinate within the SiC lattice. Empty cells correspond to missing or unreported values.}
\label{tab:summary}
\end{table*}

\clearpage
\renewcommand{\arraystretch}{1.3}
\begin{sidewaystable}
\centering
\begin{tabular}{lrrrrrrrrrrrrrrrrrrrrrrrrrrr}
\toprule
 & C1 & C2 & C3 & Si1 & Si2 & Si3 & Si4 & Si5 & Si6 & Si7 & Si8 & Si9 & Si10 & Si11 & Si12 & Si13 & Si14 & Si15 & Si17 & Si18 & Si19 & Si21 & Si22 & Si23 & Si24 & Si25 & Si26 \\
\midrule
C1 & - & - & 1.33 & - & 6.44 & - & - & - & 9.87 & - & 0.50 & 0.92 & 185.61 & - & 0.67 & 10.94 & - & - & - & - & - & - & - & - & - & - & - \\
C2 & - & - & - & - & - & - & - & - & - & 3.27 & - & - & - & 2.09 & 2.80 & - & - & - & - & - & 11.10 & - & 4.40 & - & 2.40 & - & - \\
C3 & - & - & - & - & - & - & - & - & - & - & - & - & - & - & - & - & - & - & - & - & - & - & - & - & - & - & - \\
Si1 & 2.89 & 2.46 & 0.20 & - & 7.80 & 2.28 & 1.60 & 0.50 & 2.18 & 4.16 & 0.70 & 1.25 & 4.90 & - & 2.06 & 1.78 & - & 0.80 & 1.50 & - & 1.07 & - & - & - & 1.25 & - & 0.71 \\
Si2 & 6.34 & - & - & - & - & 2.37 & 1.04 & 4.16 & 2.22 & - & - & - & 3.50 & - & - & 10.37 & 2.12 & 3.42 & 3.29 & - & - & - & - & - & - & 0.53 & 1.39 \\
Si3 & 0.66 & 0.71 & 0.98 & - & 2.33 & - & 80.06 & - & 3.95 & 0.99 & 4.57 & 0.71 & - & - & 1.14 & 0.50 & - & 2.16 & - & - & - & 1.70 & - & 5.90 & - & - & 4.39 \\
Si4 & 0.50 & 1.59 & - & - & 1.05 & 80.06 & - & 0.50 & 1.85 & 1.20 & 26.28 & 5.11 & - & - & 1.15 & 0.50 & - & 1.05 & - & - & - & 3.80 & - & 3.60 & - & - & 0.05 \\
Si5 & 0.96 & - & - & - & 4.13 & - & - & - & 0.50 & - & 0.50 & 0.50 & - & - & 0.50 & 2.11 & 79.65 & 13.39 & 35.80 & - & - & - & - & - & - & 5.10 & - \\
Si6 & 9.62 & 0.33 & 2.99 & - & 2.22 & 4.15 & 1.93 & - & - & - & 0.50 & 0.67 & - & 0.56 & 1.10 & 0.50 & - & - & - & - & - & 1.84 & - & 1.47 & - & - & 0.63 \\
Si7 & - & 3.35 & - & - & - & - & - & - & - & - & - & - & - & 2.20 & 80.80 & - & - & - & - & 4.36 & - & - & 6.00 & - & 4.60 & 0.86 & - \\
Si8 & - & - & - & - & - & - & - & - & - & - & - & - & - & - & 0.50 & - & - & - & - & - & - & 6.30 & - & 5.20 & - & - & - \\
Si9 & - & - & - & - & - & - & - & - & - & - & 4.31 & - & - & - & 0.77 & - & - & - & - & - & - & - & - & 35.20 & - & - & 3.30 \\
Si10 & - & - & - & - & - & - & - & - & - & - & - & - & - & - & - & - & - & - & - & - & - & - & - & - & - & - & - \\
Si11 & - & - & - & - & - & - & - & - & - & - & - & - & - & - & - & - & - & - & - & - & - & - & - & - & - & - & - \\
Si12 & 0.73 & - & - & - & - & - & - & - & 1.02 & - & - & - & - & 0.75 & - & - & 0.50 & 0.66 & 1.20 & - & 0.74 & - & 4.10 & - & 2.02 & - & 0.50 \\
Si13 & - & - & 7.10 & - & - & - & - & - & - & - & - & - & - & - & - & - & 1.40 & 1.92 & 2.58 & - & - & - & - & - & - & - & - \\
Si14 & - & - & - & - & - & - & - & - & - & - & - & - & - & - & - & - & - & 2.55 & 13.02 & - & - & - & - & - & - & 3.70 & - \\
Si15 & - & - & 4.30 & - & - & - & - & - & - & - & - & - & - & - & - & - & - & - & - & - & - & - & - & - & - & 4.30 & - \\
Si17 & - & - & - & - & - & - & - & - & - & - & - & - & - & - & - & - & - & - & - & - & - & - & - & - & 2.60 & 4.40 & - \\
Si18 & - & 7.20 & - & - & - & - & - & - & - & - & - & - & - & - & 4.87 & - & - & - & - & - & 2.50 & - & 18.50 & - & - & - & - \\
Si19 & - & - & - & - & - & - & - & - & - & - & - & - & - & 25.50 & - & - & - & - & - & - & - & - & - & - & 4.20 & 0.70 & - \\
Si21 & - & - & - & - & - & - & - & - & - & - & - & - & - & - & - & - & - & - & - & - & - & - & - & - & - & - & - \\
Si22 & - & 3.30 & - & - & - & - & - & - & - & - & - & - & - & - & - & - & - & - & - & - & 1.90 & - & - & - & - & - & - \\
Si23 & - & - & - & - & - & - & - & - & - & - & - & - & - & - & - & - & - & - & - & - & - & - & - & - & - & - & - \\
Si24 & - & - & - & - & - & - & - & - & - & - & - & - & - & 4.00 & - & - & - & - & - & - & - & - & - & - & - & 4.00 & - \\
Si25 & - & 0.50 & - & - & - & - & - & - & - & - & - & - & - & 4.50 & - & - & - & - & - & - & 0.60 & - & - & - & - & - & - \\
Si26 & - & - & - & - & - & - & - & - & - & - & 4.40 & - & - & - & - & - & - & 3.90 & - & - & - & 1.40 & - & 3.20 & - & 1.70 & - \\
\bottomrule
\end{tabular}
\caption{Measured couplings in the electron subdomain $m_s = -3/2$.}
\label{tab:couplings_m32}
\end{sidewaystable}
\renewcommand{\arraystretch}{1}

\renewcommand{\arraystretch}{1.3}
\begin{sidewaystable}
\begin{tabular}{lrrrrrrrrrrrrrrrrrrrrrrrrrrr}
\toprule
 & C1 & C2 & C3 & Si1 & Si2 & Si3 & Si4 & Si5 & Si6 & Si7 & Si8 & Si9 & Si10 & Si11 & Si12 & Si13 & Si14 & Si15 & Si17 & Si18 & Si19 & Si21 & Si22 & Si23 & Si24 & Si25 & Si26 \\
\midrule
C1 & - & - & - & - & - & - & - & - & - & - & - & - & - & - & - & - & - & - & - & - & - & - & - & - & - & - & - \\
C2 & - & - & - & - & - & - & - & - & - & - & - & - & - & - & - & - & - & - & - & - & - & - & - & - & - & - & - \\
C3 & - & - & - & - & - & - & - & - & - & - & - & - & - & - & - & - & - & - & - & - & - & - & - & - & - & - & - \\
Si1 & 2.72 & 2.38 & - & - & 4.66 & 2.33 & 1.46 & 0.63 & 2.05 & 4.80 & - & - & - & - & 0.88 & - & - & - & - & - & - & - & - & - & - & - & - \\
Si2 & 6.48 & - & - & 4.71 & - & 1.97 & - & 3.39 & 1.86 & - & - & - & - & - & - & - & - & - & - & - & - & - & - & - & - & - & - \\
Si3 & - & - & - & - & 1.93 & - & 79.89 & - & 3.31 & - & 3.98 & - & - & - & - & - & - & - & - & - & - & - & - & 6.02 & - & - & - \\
Si4 & - & - & - & - & - & - & - & - & - & - & 25.20 & - & - & - & - & - & - & - & - & - & - & 4.00 & - & 4.02 & - & - & - \\
Si5 & - & - & - & - & 3.53 & - & - & - & - & - & - & - & - & - & - & 1.85 & 81.37 & 13.98 & 35.71 & - & - & - & - & - & - & 5.36 & - \\
Si6 & - & - & - & - & - & - & - & - & - & - & - & - & - & - & - & - & - & 1.99 & - & - & - & - & - & - & - & - & - \\
Si7 & - & 3.50 & - & - & - & - & - & - & - & - & - & - & - & - & - & - & - & - & - & 4.30 & - & - & 6.10 & - & - & - & - \\
Si8 & - & - & - & - & - & - & - & - & - & - & - & - & - & - & - & - & - & - & - & - & - & - & - & - & - & - & - \\
Si9 & - & - & - & - & - & - & - & - & - & - & - & - & - & - & - & - & - & - & - & - & - & - & - & - & - & - & - \\
Si10 & - & - & - & - & - & - & - & - & - & - & - & - & - & - & - & - & - & - & - & - & - & - & - & - & - & - & - \\
Si11 & - & - & - & - & - & - & - & - & - & - & - & - & - & - & - & - & - & - & - & - & - & - & - & - & - & - & - \\
Si12 & - & - & - & - & - & - & - & - & - & - & - & - & - & - & - & - & - & - & - & - & - & - & - & - & - & - & - \\
Si13 & - & - & - & - & - & - & - & - & - & - & - & - & - & - & - & - & - & - & - & - & - & - & - & - & - & - & - \\
Si14 & - & - & - & - & - & - & - & - & - & - & - & - & - & - & - & - & - & - & - & - & - & - & - & - & - & - & - \\
Si15 & - & - & - & - & - & - & - & - & - & - & - & - & - & - & - & - & - & - & - & - & - & - & - & - & - & - & - \\
Si17 & - & - & - & - & - & - & - & - & - & - & - & - & - & - & - & - & - & - & - & - & - & - & - & - & - & - & - \\
Si18 & - & - & - & - & - & - & - & - & - & - & - & - & - & - & - & - & - & - & - & - & - & - & - & - & - & - & - \\
Si19 & - & - & - & - & - & - & - & - & - & - & - & - & - & - & - & - & - & - & - & - & - & - & - & - & - & - & - \\
Si21 & - & - & - & - & - & - & - & - & - & - & - & - & - & - & - & - & - & - & - & - & - & - & - & - & - & - & - \\
Si22 & - & - & - & - & - & - & - & - & - & - & - & - & - & - & - & - & - & - & - & - & - & - & - & - & - & - & - \\
Si23 & - & - & - & - & - & - & - & - & - & - & - & - & - & - & - & - & - & - & - & - & - & - & - & - & - & - & - \\
Si24 & - & - & - & - & - & - & - & - & - & - & - & - & - & - & - & - & - & - & - & - & - & - & - & - & - & - & - \\
Si25 & - & - & - & - & - & - & - & - & - & - & - & - & - & - & - & - & - & - & - & - & - & - & - & - & - & - & - \\
Si26 & - & - & - & - & - & - & - & - & - & - & - & - & - & - & - & - & - & - & - & - & - & - & - & - & - & - & - \\
\bottomrule
\end{tabular}

\caption{Measured couplings in the electron subdomain $m_s = +3/2$.}
\label{tab:couplings_p32}
\end{sidewaystable}
\renewcommand{\arraystretch}{1}

\begin{sidewaystable}
\renewcommand{\arraystretch}{1.4}
\centering
\begin{tabular}{lrrrrrrrrrrrrrrrrrrrrrrrrrrr}
\toprule
 & C1 & C2 & C3 & Si1 & Si2 & Si3 & Si4 & Si5 & Si6 & Si7 & Si8 & Si9 & Si10 & Si11 & Si12 & Si13 & Si14 & Si15 & Si17 & Si18 & Si19 & Si21 & Si22 & Si23 & Si24 & Si25 & Si26 \\
\midrule
C1 & - & - & 1.33 & 2.81 & 6.44 & 0.66 & 0.50 & 0.96 & 9.74 & - & 0.50 & 0.92 & 185.61 & - & 0.70 & 10.94 & - & - & - & - & - & - & - & - & - & - & - \\
C2 & - & - & - & 2.42 & - & 0.71 & 1.59 & - & 0.33 & 3.41 & - & - & - & 2.09 & 2.80 & - & - & - & - & 7.20 & 11.10 & - & 3.85 & - & 2.40 & 0.50 & - \\
C3 & 1.33 & - & - & 0.20 & - & 0.98 & - & - & 2.99 & - & - & - & - & - & - & 7.10 & - & 4.30 & - & - & - & - & - & - & - & - & - \\
Si1 & 2.81 & 2.42 & 0.20 & - & 6.25 & 2.30 & 1.53 & 0.56 & 2.12 & 4.48 & 0.70 & 1.25 & 4.90 & - & 1.47 & 1.78 & - & 0.80 & 1.50 & - & 1.07 & - & - & - & 1.25 & - & 0.71 \\
Si2 & 6.44 & - & - & 6.25 & - & 2.15 & 1.04 & 3.80 & 2.04 & - & - & - & 3.50 & - & - & 10.37 & 2.12 & 3.42 & 3.29 & - & - & - & - & - & - & 0.53 & 1.39 \\
Si3 & 0.66 & 0.71 & 0.98 & 2.30 & 2.15 & - & 79.97 & - & 3.68 & 0.99 & 4.28 & 0.71 & - & - & 1.14 & 0.50 & - & 2.16 & - & - & - & 1.70 & - & 5.96 & - & - & 4.39 \\
Si4 & 0.50 & 1.59 & - & 1.53 & 1.04 & 79.97 & - & 0.50 & 1.89 & 1.20 & 25.74 & 5.11 & - & - & 1.15 & 0.50 & - & 1.05 & - & - & - & 3.90 & - & 3.81 & - & - & 0.05 \\
Si5 & 0.96 & - & - & 0.56 & 3.80 & - & 0.50 & - & 0.50 & - & 0.50 & 0.50 & - & - & 0.50 & 1.98 & 80.51 & 13.69 & 35.75 & - & - & - & - & - & - & 5.23 & - \\
Si6 & 9.74 & 0.33 & 2.99 & 2.12 & 2.04 & 3.68 & 1.89 & 0.50 & - & - & 0.50 & 0.67 & - & 0.56 & 1.06 & 0.50 & - & 1.99 & - & - & - & 1.84 & - & 1.47 & - & - & 0.63 \\
Si7 & - & 3.41 & - & 4.48 & - & 0.99 & 1.20 & - & - & - & - & - & - & 2.20 & 80.80 & - & - & - & - & 4.33 & - & - & 6.05 & - & 4.60 & 0.86 & - \\
Si8 & 0.50 & - & - & 0.70 & - & 4.28 & 25.74 & 0.50 & 0.50 & - & - & 4.31 & - & - & 0.50 & - & - & - & - & - & - & 6.30 & - & 5.20 & - & - & 4.40 \\
Si9 & 0.92 & - & - & 1.25 & - & 0.71 & 5.11 & 0.50 & 0.67 & - & 4.31 & - & - & - & 0.77 & - & - & - & - & - & - & - & - & 35.20 & - & - & 3.30 \\
Si10 & 185.61 & - & - & 4.90 & 3.50 & - & - & - & - & - & - & - & - & - & - & - & - & - & - & - & - & - & - & - & - & - & - \\
Si11 & - & 2.09 & - & - & - & - & - & - & 0.56 & 2.20 & - & - & - & - & 0.75 & - & - & - & - & - & 25.50 & - & - & - & 4.00 & 4.50 & - \\
Si12 & 0.70 & 2.80 & - & 1.47 & - & 1.14 & 1.15 & 0.50 & 1.06 & 80.80 & 0.50 & 0.77 & - & 0.75 & - & - & 0.50 & 0.66 & 1.20 & 4.87 & 0.74 & - & 4.10 & - & 2.02 & - & 0.50 \\
Si13 & 10.94 & - & 7.10 & 1.78 & 10.37 & 0.50 & 0.50 & 1.98 & 0.50 & - & - & - & - & - & - & - & 1.40 & 1.92 & 2.58 & - & - & - & - & - & - & - & - \\
Si14 & - & - & - & - & 2.12 & - & - & 80.51 & - & - & - & - & - & - & 0.50 & 1.40 & - & 2.55 & 13.02 & - & - & - & - & - & - & 3.70 & - \\
Si15 & - & - & 4.30 & 0.80 & 3.42 & 2.16 & 1.05 & 13.69 & 1.99 & - & - & - & - & - & 0.66 & 1.92 & 2.55 & - & - & - & - & - & - & - & - & 4.30 & 3.90 \\
Si17 & - & - & - & 1.50 & 3.29 & - & - & 35.75 & - & - & - & - & - & - & 1.20 & 2.58 & 13.02 & - & - & - & - & - & - & - & 2.60 & 4.40 & - \\
Si18 & - & 7.20 & - & - & - & - & - & - & - & 4.33 & - & - & - & - & 4.87 & - & - & - & - & - & 2.50 & - & 18.50 & - & - & - & - \\
Si19 & - & 11.10 & - & 1.07 & - & - & - & - & - & - & - & - & - & 25.50 & 0.74 & - & - & - & - & 2.50 & - & - & 1.90 & - & 4.20 & 0.65 & - \\
Si21 & - & - & - & - & - & 1.70 & 3.90 & - & 1.84 & - & 6.30 & - & - & - & - & - & - & - & - & - & - & - & - & - & - & - & 1.40 \\
Si22 & - & 3.85 & - & - & - & - & - & - & - & 6.05 & - & - & - & - & 4.10 & - & - & - & - & 18.50 & 1.90 & - & - & - & - & - & - \\
Si23 & - & - & - & - & - & 5.96 & 3.81 & - & 1.47 & - & 5.20 & 35.20 & - & - & - & - & - & - & - & - & - & - & - & - & - & - & 3.20 \\
Si24 & - & 2.40 & - & 1.25 & - & - & - & - & - & 4.60 & - & - & - & 4.00 & 2.02 & - & - & - & 2.60 & - & 4.20 & - & - & - & - & 4.00 & - \\
Si25 & - & 0.50 & - & - & 0.53 & - & - & 5.23 & - & 0.86 & - & - & - & 4.50 & - & - & 3.70 & 4.30 & 4.40 & - & 0.65 & - & - & - & 4.00 & - & 1.70 \\
Si26 & - & - & - & 0.71 & 1.39 & 4.39 & 0.05 & - & 0.63 & - & 4.40 & 3.30 & - & - & 0.50 & - & - & 3.90 & - & - & - & 1.40 & - & 3.20 & - & 1.70 & - \\
\bottomrule
\end{tabular}

\caption{Couplings averaged over the two electron subdomains $m_s = \pm3/2$. Here, first an average within each subspace is performed (e.g. average over Si2--Si3 and Si3--Si2) and subsequently over both subspaces Si2--Si3 (-3/2) and Si2--Si3 (+3/2). }
\label{tab:couplings_avg}
\end{sidewaystable}
\renewcommand{\arraystretch}{1}
\newpage
\begin{algorithm}
\SetAlgoLined
\SetKwFunction{FMain}{iterative\_spin\_placement}
\SetKwProg{Fn}{Function}{:}{}

\Fn{\FMain{spins, tolerance}}{

    remaining\_spins $\leftarrow$ copy of spins\;

    Create initial solution as OrderedDict:\;
    \Indp place Si1 at its known reference position\;
    \Indm

    solutions $\leftarrow$ [(initial\_solution, residual = 0)]\;

    Remove Si1 from remaining\_spins\;

    \BlankLine
    \tcp{MAIN LOOP: place spins sequentially}
    \ForEach{spin\_name in remaining\_spins}{

        Determine spin\_gamma (13C or 29Si)\;

        Select lattice\_sites (Si or C lattice)\;

        new\_solutions $\leftarrow$ empty list\;

        \BlankLine
        \tcp{Loop over existing candidate solutions}
        \ForEach{(placed\_spins, current\_residual) in solutions}{

            Sort placed\_spins descending by $|$coupling(ref, new spin)$|$\;

            best\_matches $\leftarrow$ None\;

            \BlankLine
            \tcp{Loop over each reference spin already placed}
            \ForEach{ref\_spin in placed\_spins}{

                expected\_coupling $\leftarrow$ measured coupling(ref\_spin, spin\_name)\;

                \If{expected\_coupling invalid}{continue\;}

                Adjust tolerance heuristically\;

                \BlankLine
                \tcp{FIRST reference spin: evaluate all candidate lattice positions}
                \uIf{best\_matches is None}{

                    coords $\leftarrow$ all lattice positions\;

                    $r\_{\text{vecs}} \leftarrow \text{coords} - \text{ref\_position}$\;

                    $C\_{ij} \leftarrow \text{dipolar\_coupling\_vectorized}(r\_{\text{vecs}})$\;

                    errors $\leftarrow |$expected\_coupling $- C\_{ij}|$\;

                    mask $\leftarrow$ errors $<$ tolerance\;

                    best\_matches $\leftarrow$ structured array storing:\;
                    \Indp coords[mask] with spin\_gamma appended\;
                    residuals $=$ errors[mask]\;
                    \Indm
                }
                \BlankLine
                \tcp{SUBSEQUENT reference spins: filter only previously valid coords}
                \uElse{

                    $r\_{\text{vecs}} \leftarrow$ best\_matches.coords $-$ ref\_position\;

                    $C\_{ij} \leftarrow \text{dipolar\_coupling\_vectorized}(r\_{\text{vecs}})$\;

                    errors $\leftarrow |$expected\_coupling $- C\_{ij}|$\;

                    mask $\leftarrow$ errors $<$ tolerance\;

                    best\_matches $\leftarrow$ best\_matches[mask]\;

                    Update residuals accordingly\;
                }

            }

            \BlankLine
            \tcp{After checking all reference spins}
            \If{best\_matches not empty}{
                Sort best\_matches by residual (ascending)\;

                \ForEach{match in best\_matches}{
                    new\_placed $\leftarrow$ copy placed\_spins\;
                    Add spin\_name with match.coord\;

                    residual\_sum $\leftarrow$ compute\_residual\_sum(new\_placed)\;

                    Append (new\_placed, residual\_sum) to new\_solutions\;
                }
            }

        }

        \BlankLine
        \tcp{Update solution pool}
        \If{new\_solutions empty}{raise warning\;}

        Sort new\_solutions by residual (ascending)\;

        Save solutions to disk with filename based on spin\_name\;

        solutions $\leftarrow$ new\_solutions\;
    }

    \BlankLine
    \Return{all solutions sorted by residual}\;

}
\end{algorithm}

\bibliographystyle{apsrev4-2}
\bibliography{bibliography}